\documentstyle[12pt]{article}
\makeatletter
\@addtoreset{equation}{section}
\makeatother

\newcommand{\ret}{\nonumber\\}

\newcommand{\lesssim}
{\mathrel{\raisebox{-2.8pt}{\mbox{$\stackrel{\textstyle <}{\sim}$}}}}

\newtheorem{theorem}{Theorem}
\newtheorem{lemma}[theorem]{Lemma}

\newcommand{\abs}[1]{\left|#1\right|}

\newcommand{\normi}[1]{\left\Vert#1\right\Vert_{\infty}}
\newcommand{\bktt}[1]{\bkt{#1}_{t}}
\newcommand{\rbk}[1]{\left(#1\right)}
\newcommand{\sqbk}[1]{\left[#1\right]}
\newcommand{\cbk}[1]{\left\{#1\right\}}
\newcommand{\bkt}[1]{\left\langle#1\right\rangle}

\newcommand{\calH}{{\cal H}}
\newcommand{\HiS}{\calH_{\rm S}}
\newcommand{\HiB}{\calH_{\rm B}}
\newcommand{\ep}{\varepsilon}
\newcommand{\tE}{\widetilde{E}}
\newcommand{\HS}{H_{\rm S}}
\newcommand{\HB}{H_{\rm B}}
\newcommand{\one}{{\bf 1}}

\newcommand{\oneB}{{\bf 1}_{\rm B}}
\newcommand{\can}[2]{\bkt{#1}^{\rm can}_{#2}}

\renewcommand{\phi}{\varphi}
\newcommand{\gbar}{\bar{\gamma}}
\newcommand{\proof}{\par\bigskip\noindent{\em Proof:\ }}
\newcommand{\qed}{~\rule{2mm}{2mm}\par\bigskip}
\newcommand{\Sch}{Schr\"{o}dinger}
\newcommand{\Bk}{B_{k}}
\newcommand{\epj}{\varepsilon_{j}}
\newcommand{\epn}{\varepsilon_{n}}
\newcommand{\lra}{\leftrightarrow}
\newcommand{\tb}{\tilde{b}}
\newcommand{\tM}{\widetilde{M}}
\newcommand{\tl}{\tilde{\ell}}
\newcommand{\bb}{\bar{b}}
\newcommand{\bu}{\bar{u}}
\newcommand{\bal}{\overline{\alpha}}
\newcommand{\tU}{\widetilde{U}}
\newcommand{\ellt}{\ell_{\rm t}}
\newcommand{\sT}{{\sf T}}
\newcommand{\sM}{{\sf M}}
\newcommand{\sD}{{\sf D}}
\newcommand{\eeD}[1]{\rbk{\matrix{e_{#1}&0\cr0&\overline{e_{#1}}}}}
\newcommand{\bG}{\overline{G}}
\newcommand{\hI}{\hat{I}}
\newcommand{\hell}{\hat{\ell}}
\newcommand{\tzeta}{\tilde{\zeta}}
\newcommand{\tZ}{\widetilde{Z}}
\newcommand{\txi}{\tilde{\xi}}
\newcommand{\kmax}{k_{\rm max}}
\newcommand{\kmin}{k_{\rm min}}
\newcommand{\bas}{\bar{s}}
\newcommand{\teta}{\tilde{\eta}}
\begin{document}
\begin{flushright}
{\footnotesize
Technical note (not for publication as it is)
}
\end{flushright}
\begin{center}
{\large\bf
Note on
``From Quantum Dynamics to the Canonical Distribution
-- A Rigorous Derivation in Special Models''\\
}
\bigskip
Hal Tasaki\footnote{
Department of Physics,
Gakushuin University,
Mejiro, Toshima-ku, Tokyo 171,
JAPAN

electronic address: hal.tasaki@gakushuin.ac.jp
}
\end{center}
\par\bigskip\bigskip
\tableofcontents
\newpage
\section{About this note}
This is a note associated with my paper
``From Quantum Dynamics to the Canonical Distribution
-- A Rigorous Derivation in Special Models''
(cond-mat/9707253).
Here I describe all the technical details which are not discussed in
the main paper.

Please note that this is not (yet) written as a regular paper.
I did not include any introductory materials or physical discussions.
The proofs may not be optimally organized yet.

The preset note is organized as follows.
In Section~\ref{s:nonres}, we prove a simple claim about the
robustness of the non-resonance condition that we mentioned in the
main paper.
In Section~\ref{s:theorem}, we prove the Theorem in the main paper.
The theorem is essentially an application of the Chebysehv's
inequality, and the proof is easy.
In Section~\ref{s:lemma}, we prove the Lemma in the main paper.
As is clear from the table of contents, this is the hardest and the
most technical part in our analysis.
We have summarized the basic strategy in the beginning of the section.
Section~\ref{s:gen} is independent from the rigorous example we
discuss in the main paper and in the present note.
Here we deal with much more general class of models, and show that
the ``half'' of the ``hypothesis of equal weights for eigenstates''
can be proved rather easily.

\section{Non-resonance condition}
\label{s:nonres}
Let us prove the statement about robustness of the
non-resonance condition mentioned in the footnote [8] of the main
paper.

Let $E_{i}$ be an eigenvalue of $H$ and let the corresponding
normalized eigenstate be
\begin{equation}
	\Phi_{E_{i}}=
	\sum_{j=1}^{n}\sum_{k=1}^{N}
	\phi_{(j,k)}^{(i)}\Psi_{j}\otimes\Gamma_{k}.
	\label{PhiEi}
\end{equation}
From the first order perturbation theory, we get
\begin{equation}
	\frac{\partial E_{i}}{\partial B_{k'}}
	=
	\sum_{j=1}^{n}|\phi_{(j,k')}^{(i)}|^{2}.
	\label{dEdB}
\end{equation}

Suppose that the energy spectrum $\{E_{i}\}$ violates the
non-resonance condition for some $E_{i}$'s in the range
$E_{i}\ge\ep_{n}+2\lambda$.
More precisely, we assume that there are $i_1,i_{2},i_{3},i_{4}$ such
that
$R=E_{i_{1}}-E_{i_{2}}-(E_{i_{3}}-E_{i_{4}})=0$
holds.

We now shift all the $B_{k'}$'s in the lowest band $(0,\delta)$ by a
small amount, say $d$, keeping their spacing unchanged.
If
\begin{equation}
	\frac{\partial R}{\partial d}=
	\frac{\partial E_{1}}{\partial d}
	-\frac{\partial E_{2}}{\partial d}
	-\frac{\partial E_{3}}{\partial d}
	+\frac{\partial E_{4}}{\partial d}
	\label{dR/dd}
\end{equation}
is nonvanishing, then we can conclude that the resonance is lifted for
any small shift $d$.
Also note that
no new resonances are generated if we keep $d$
sufficiently small.

Let us assume that $\frac{\partial R}{\partial d}$ happens to
be vanishing.
In such a (extremely rare) situation,
we shift all the $B_{k'}$'s in the second band $(\delta,2\delta)$
by  $d'$.
We can then say that
\begin{equation}
	\frac{\partial R}{\partial d'}=
	\frac{\partial E_{1}}{\partial d'}
	-\frac{\partial E_{2}}{\partial d'}
	-\frac{\partial E_{3}}{\partial d'}
	+\frac{\partial E_{4}}{\partial d'}
	\label{dR/dd'}
\end{equation}
is nonvanishing.
To see this we recall the representation (\ref{dEdB}).
That $\frac{\partial R}{\partial d}=0$ means that there is a very
special relation between
$\sum_{k';{\rm lowest}}\sum_{j=1}^n|\phi^{(i_{\mu})}_{(j,k')}|^2$
with
$\mu=1,2,3,4$.
Since $\phi_{(j,k)}$ is determined as the solution of the \Sch\
equation (4) in the main paper, it has different decay properties in
the ``classically inaccessible regions'' for different values of $E$.
This means that
$\sum_{k';{\rm second}}\sum_{j=1}^n|\phi^{(i_{\mu})}_{(j,k')}|^2$
with $\mu=1,2,3,4$ cannot satisfy the same special relation as the
corresponding quantities of the lowest band.
So we conclude that the resonance is lifted by a small $d'$.

The same argument works for the cases with multiple resonances.
We see that all the resonances go away if we allow small
$d$ and $d'$.

\section{Proof of Theorem}
\label{s:theorem}
Let us prove the Theorem in the main paper.
We first state and prove a general lemma, on which the desired theorem
relies.
Let $\Phi_{E'}$ be the eigenstate of the total Hamiltonian $H$ with
the eigenvalue $E'$.
We assume that the initial state $\Phi(0)$ of the system is expanded
 as
 \begin{equation}
 	\Phi(0)=\sum_{E'}\gamma_{E'}\Phi_{E'},
 	\label{Phi0exp}
 \end{equation}
 and set
 \begin{equation}
 	\gbar=\max_{E'}|\gamma_{E'}|.
 	\label{gbar}
 \end{equation}
 As in the main paper $\bkt{A}_{t}$ denote the expectation value of
 the operator $A$ of the subsystem in the state at time $t$.
 We have shown in the main paper that
 \begin{equation}
 	\overline{\cbk{\bkt{A}_{t}-\overline{\bkt{A}_{t}}}^{2}}
 	\le n^{2}(\normi{A})^{2}\gbar^{2}.
 	\label{A-A<}
 \end{equation}
 Then we have
\begin{lemma}
\label{l:gen}
Let $A$ be an arbitrary operator of the subsystem.
Let $\kappa>0$ and $\Delta>0$ be arbitrary constants.
Then there exists a ($\kappa$-dependent) constant $T>0$, and we have
\begin{equation}
	\frac{\tau_{\Delta}(T)}{T}
	\le
	\frac{(1+\kappa)n^{2}(\normi{A})^{2}\gbar^{2}}{\Delta^{2}},
	\label{tau/T}
\end{equation}
where $\tau_{\Delta}(T)$ is the total length of the intervals
within  $0\le t\le T$ at which
\begin{equation}
	|\bkt{A}_{t}-\overline{\bkt{A}_{t}}|
	\ge
	\Delta
	\label{ADelta}
\end{equation}
holds.
\end{lemma}
\proof
This is nothing but the Chebyshev's inequality, but we give a proof
for completeness.
For a function $f(t)$ of $t$, we let
\begin{equation}
	v_{T}[f(t)]=\frac{1}{T}
	\int_{0}^{T}dt(f(t)-\overline{f(t)})^{2}.
	\label{vtf}
\end{equation}
Since (\ref{A-A<}) implies
\begin{equation}
	\lim_{T\uparrow\infty}
	v_{T}[\bkt{A}_{t}]\le n^{2}(\normi{A})^{2}\gbar^{2},
	\label{limvT}
\end{equation}
we see from continuity that for a given $\kappa>0$, there is $T>0$
such that
\begin{equation}
	v_{T}[\bkt{A}_{t}]\le
	(1+\kappa)n^{2}(\normi{A})^{2}\gbar^{2}.
	\label{vT<}
\end{equation}

Now observe that
\begin{equation}
	\chi\sqbk{\abs{\bktt{A}-\overline{\bktt{A}}}\ge\Delta}
	\le
	\frac{\rbk{\bktt{A}-\overline{\bktt{A}}}^2}
	{\Delta^2},
	\label{chi<}
\end{equation}
where the characteristic function $\chi$ is defined by
$\chi[\mbox{true}]=1$ and $\chi[\mbox{false}]=0$.
By averaging (\ref{chi<}) over $t$ such that $0\le t\le T$, we find
\begin{equation}
	\tau_{\Delta}(T)\le\frac{v_{T}[\bktt{A}]}{\Delta^2},
	\label{tauDelta2}
\end{equation}
which with (\ref{vT<}) implies the desired (\ref{tau/T}).\qed

In order to prove the Theorem in the main paper, we have to
evaluate $\gbar$ and choose appropriate $\Delta$.

We recall that we have the initial state of the form
\begin{equation}
	\Phi(0)=\Psi_{n}\otimes\sum_{k}\alpha_{k}\Gamma_{k},
	\label{Phi00}
\end{equation}
with $\alpha_{k}$ nonvanishing only for $k$ such that
\begin{equation}
	|E-(\ep_{n}+B_{k})|\le \frac{\ep_{n}}{2},
	\label{alphak}
\end{equation}
for a fixed $E$ such that
$\epn+3\lambda\le E\le B_{\rm max}-3\lambda$.
To evaluate $\gbar$, we note that
\begin{equation}
	\gamma_{E'}=\bkt{\Phi_{E'},\Phi(0)}
	=\sum_{k}\overline{\phi_{(n,k)}}\alpha_{k},
	\label{gE'}
\end{equation}
where we wrote
\begin{equation}
	\Phi_{E'}=
	\sum_{j=1}^n\sum_{k=1}^N
	\phi_{(j,k)}\Psi_{j}\otimes\Gamma_{k}.
	\label{PhiE'}
\end{equation}
We shall prove in Section~\ref{s:decay} that $|\phi_{(n,k)}|$ is less
than $O(\exp[-{\rm const.}L^{1/3}])$ outside the interval
$\{k_{\rm min},\ldots,k_{\rm max}\}$ determined by the condition
\begin{equation}
	|E'-(\epn+B_{j})|\le \lambda+{\rm const.}L^{-1/3}.
	\label{phik}
\end{equation}

For $E'$ such that $|E'-E|\ge\lambda+(\ep_{n}/2)$, the two ranges
(\ref{alphak}), (\ref{phik}) have no overlaps, and we see that
$|\gamma_{E'}|$ is small.

So we assume $|E'-E|\le\lambda+(\ep_{n}/2)$.
From (\ref{gE'}), we have
\begin{equation}
	|\gamma_{E'}|\le(\max_{k}|\alpha_{k}|)\sum_{k}|\phi_{(n,k)}|.
	\label{|gE'|}
\end{equation}
By using the above observation about $|\phi_{(n,k)}|$, we find
\begin{eqnarray}
	\sum_{k}|\phi_{(n,k)}|
	& \le &
	\sum_{k=k_{\rm min}}^{k_{\rm max}}|\phi_{(n,k)}|
	+{\rm const.}\exp[-{\rm const.}L^{1/3}]
	\ret
	 & \le &
	\cbk{
	\rbk{\sum_{k}|\phi_{(n,k)}|^2}
	\rbk{\sum_{k=k_{\rm min}}^{k_{\rm max}}1}
	}^{1/2}
	+{\rm const.}\exp[-{\rm const.}L^{1/3}]
	\ret
	 & \le &
	(k_{\rm max}-k_{\rm min})^{1/2}
	+{\rm const.}\exp[-{\rm const.}L^{1/3}]
	\ret
	 & \le &
	\sqrt{2\lambda\rho(E'-\ep_{n}+\lambda)}
	+{\rm const.}\exp[-{\rm const.}L^{1/3}]
	\ret
	 & \le &
	\sqrt{2\lambda\rho(E)}.
	\label{philong}
\end{eqnarray}
By using the assume bound for $\alpha_{k}$ (see the main paper), we
find that
\begin{equation}
	|\gamma_{E'}|^2\le\frac{2c'\lambda}{\ep_{n}},
	\label{gE'<}
\end{equation}
and can set $\gbar^{2}={2c'\lambda}/{\ep_{n}}$.

We recall that the desired theorem in the main paper does not specify
the operator $A$.
By linearity we can replace the phrase ``for any operator $A$'' in
the Theorem by ``for any of the $n^2$ operators
$A_{\mu,\nu}$ defined by
$(A_{\mu,\nu})_{j,j'}=\delta_{\mu,j}\delta_{\nu,j'}$.''

We now let $A$ be one of the $A_{\mu,\nu}$'s, and apply
Lemma~\ref{l:gen} by setting
\begin{equation}
	\Delta=n^2\normi{A}\rbk{\frac{\lambda}{\ep_{n}}}^{1/3},
	\label{Deltavalue}
\end{equation}
and $\kappa=1/2$.
Then we have
\begin{equation}
	\frac{\tau_{\Delta}(T)}{T}
	\le
	\frac{3}{2n^2}\rbk{\frac{\lambda}{\ep_{n}}}^{-2/3}\gbar^{2}
	=
	\frac{3c'}{n^2}\rbk{\frac{\lambda}{\ep_{n}}}^{1/3}.
	\label{tau/T2}
\end{equation}
Since each of $A_{\mu,\nu}$ can have its own ``bad'' interval, the
total length of the ``bad'' intervals of all  $A_{\mu,\nu}$ is
bounded by $n^2$ times the right-hand side of (\ref{tau/T2}).
This is what appears in the right-hand side of (15) in the main
paper.

It only remains to estimate the systematic difference between the
desired $\can{A}{\beta}$ and $\overline{\bktt{A}}$.
We first note that
\begin{eqnarray}
	\abs{
	\can{A}{\beta(E')}-\can{A}{\beta(E)}
	} & = &
	\abs{
	\int_{E}^{E'}dF\frac{d}{dE}\can{A}{\beta(F)}
	}
	\ret
	 & = &
	\abs{
	\int_{E}^{E'}dF\frac{d\beta(F)}{dE}
	\frac{d}{d\beta}\can{A}{\beta(F)}
	}
	\ret
	 & \le &
	\gamma
	\abs{
	\int_{E}^{E'}dF\rbk{
	\can{A\ep}{\beta(F)}-\can{A}{\beta(F)}\can{\ep}{\beta(F)}
	}
	}
	\ret
	 & \le &
	2\gamma|E'-E|\normi{A}\ep_{n}
	\ret
	 & \le &
	2\normi{A}\gamma\rbk{\lambda+\frac{\ep_{n}}{2}}\ep_{n}
	\ret
	 & \le &
	2\normi{A}\gamma(\epn)^2,
	\label{A-Along}
\end{eqnarray}
where $\gamma=d\beta(E+\lambda)/dE$.
Thus we see
\begin{equation}
	\abs{
	\can{A}{\beta(E)}
	-
	\sum_{E'}|\gamma_{E'}|^2\can{A}{\beta(E')}
	}
	\le
	2\normi{A}\gamma(\epn)^2.
	\label{A-Afinal}
\end{equation}
By adding this new systematic error to the error $\sigma\normi{A}$ in
the (7) of the main paper, we finally get the statement of the
Theorem.

\section{Proof of Lemma}
\label{s:lemma}
In the present rather lengthy section, we shall prove the Lemma in
the main paper, which is our main estimate.
Throughout the present section, $c_{1},c_{2},\ldots,c_{23}$
denote positive
constants which do not depend on $L$ but may depend on
$\{\ep_{j}\}$, $\lambda$, $\delta$, and $\{N_{r}\}$.

Let us give an outline of the present section.
In Section~\ref{s:proof}, we prove the statement of the Lemma, but
assuming some new lemmas which will be proved in the latter
sections.

In Section~\ref{s:reg}, we introduce the notion of ``regular
interval'', and study how the two index systems $\ell$ and $(j,k)$
are related with each other.
This is essential in getting the desired Boltzmann weights.

In Section~\ref{s:int}, we decompose the whole region
$\{1,2,\ldots,nN\}$ for the index $\ell$ into many subintervals.

The next three sections are devoted to the estimate of the
solution of the \Sch\ equation (\ref{Sch}) in each of the above
intervals.
Therefore the topics of these three sections are asymptotic
analyses of a discrete \Sch\ equation, and are not quite specific to
the problem of deriving the canonical distribution from quantum
dynamics.
Although the techniques I use are not quite original,
I present all the estimates since I could not find
necessary estimates in the literature.
In Section~\ref{s:discap}, we treat the solution in the ``classically
accessible region''.
The approximate solution is obtained by a speculation based on the
quasi-classical analysis, and the difference from the true
solution is rigorously controlled by the standard machinery of
transfer matrices.
In Section~\ref{s:contap}, we treat the solution near the ``turning
points'' where the quasi-classical analysis no longer works.
We use the solution of (rescaled) continuous \Sch\ equation as an
approximate solution, and control the difference from the true
solution inductively.
In Section~\ref{s:decay}, we control the solution in the ``classically
inaccessible regions''.

In the next two sections, we extract information about the
Boltzmann factor from the controlled approximate solutions of the
\Sch\ equation.
In Section~\ref{s:res}, we treat the ``classically
accessible region''.
We encounter an annoying phenomenon of ``resonance'', which locally
inhibits the wave function to generate the desired Boltzmann factor.
We suspect that this phenomenon is of essential character.
In Section~\ref{s:long}, we get the Boltzmann factor in the region
where the wave length of the wave function is long.
There are no resonances, and the proof is easy.

In Section~\ref{s:exp}, we fix some exponents, and complete the
lengthy proof.

\subsection{Proof}
\label{s:proof}
We write the \Sch\ equation (4) in the main paper as
\begin{equation}
	\phi_{\ell-1}+\phi_{\ell+1}+2\alpha_{\ell}\phi_{\ell}=0,
	\label{Sch}
\end{equation}
with
\begin{equation}
	\alpha_{\ell}=\frac{U_{\ell}-E}{\lambda},
	\label{alphal}
\end{equation}
where $E$ is an eigenvalue such that
\begin{equation}
	\epn+2\lambda\le E\le B_{\rm max}-2\lambda.
	\label{<E<}
\end{equation}

We shall decompose the whole range of the index $\ell$ into a disjoint
union of $\Omega$ intervals as
\begin{equation}
	\{1,2,\ldots,nN\}=
	\bigcup_{\omega=1}^\Omega I_{\omega},
	\label{dec}
\end{equation}
where the intervals $I_{\omega}$ will be specified later in
Section~\ref{s:int}.
We here note that $I_{1}$ and $I_{\Omega}$ are special intervals
which consist of the ``classically inaccessible regions'' with
$\alpha_{\ell}\le-1$ and $\alpha_{\ell}\ge1$, respectively, plus
small ranges in the ``classically accessible region'' attached to
them.
All the other intervals $I_{2},I_{3},\ldots,I_{\Omega-1}$ are in the
``classically accessible region'' $-1<\alpha_{\ell}<1$.

We recall that we have two different index systems, namely, $\ell$
with $\ell=1,2,\ldots,nN$, and $(j,k)$ with $j=1,2,\ldots,n$ and
$k=1,2,\ldots,N$, and these two are in one-to-one correspondence
$\ell\leftrightarrow(j,k)$.
We make the correspondence manifest by writing
\begin{equation}
	\ell\leftrightarrow(j(\ell),k(\ell)),
	\quad\mbox{or}\quad
	\ell(j,k)\leftrightarrow(j,k).
	\label{elljk}
\end{equation}
This is a slight abuse of notation, but we hope there will be no
confusions.

The following is our main estimate.
It sates that $\phi_{\ell}$ produces the desired Boltzmann factor in
each interval.
\begin{lemma}
\label{l:part}
Let $\phi_{\ell}$ be a normalized solution of (\ref{Sch}).
For each $\omega=1,2,\ldots,\Omega$ and each $j=1,2,\ldots,n$, we have
\begin{equation}
	\abs{
	\frac{
	\sum_{\ell\in I_{\omega}}\chi[j(\ell)=j]\abs{\phi_{\ell}}^2
	}{
	\sum_{\ell\in I_{\omega}}\abs{\phi_{\ell}}^2
	}
	-W_{j}^{(\omega)}
	}
	\le
	c_{1}W_{j}^{(\omega)}L^{-\eta},
	\label{partB}
\end{equation}
where the indicator function is defined by
$\chi[\mbox{\em true}]=1$ and
$\chi[\mbox{\em false}]=0$.
The exponent $\eta>0$ will be determined later in
Section~\ref{s:exp} (to be $1/12$).
Here
\begin{equation}
	W_{j}^{(\omega)}=
	\frac{\rho(U_{\tilde{\ell}(\omega)}-\ep_{j})}
	{\sum_{j'=1}^n
	\rho(U_{\tilde{\ell}(\omega)}-\ep_{j'})}
	\label{Wwj}
\end{equation}
is essentially the Boltzmann factor.
The index $\tilde{\ell}(\omega)$ are taken from the interval
$I_{\omega}$.
For the intervals $I_{1}$ and $I_{\Omega}$, we take the corresponding
$\tilde{\ell}(1)$ and $\tilde{\ell}(\Omega)$ from the ``classically
accessible parts'' of the intervals.
\end{lemma}
This lemma will be proved in subsequent
sections by constructing local approximations for $\phi_{\ell}$.

Given this lemma,we immediately get
\begin{lemma}
\label{l:whole}
Let $\phi_{\ell}$ be a normalized solution of (\ref{Sch}).
For each $j=1,2,\ldots,n$, we have
\begin{equation}
	\abs{
	\sum_{k=1}^N\abs{\phi_{(j,k)}}^2-W_{j}
	}
	\le
	c_{1}W_{j}L^{-\eta},
	\label{wholeB1}
\end{equation}
with $W_{j}$ satisfying
\begin{equation}
	\sum_{j=1}^nW_{j}=1,
	\label{Wnorm}
\end{equation}
and
\begin{equation}
	\frac{\rho(E-\ep_{j}-\lambda)}
	{\sum_{j'=1}^n\rho(E-\ep_{j'}+\lambda)}
	\le
	W_{j}
	\le
	\frac{\rho(E-\ep_{j}+\lambda)}
	{\sum_{j'=1}^n\rho(E-\ep_{j'}-\lambda)}.
	\label{Wbound}
\end{equation}
\end{lemma}

\proof
This is almost trivial.
By writing
$P_{\omega}=\sum_{\ell\in I_{\omega}}|\phi_{\ell}|^2$,
we have
\begin{eqnarray}
	\sum_{k=1}^N\abs{\phi_{(j,k)}}^2
	 & = &
	\sum_{\ell=1}^{nN}
	\chi[j(\ell)=j]|\phi_{\ell}|^2
	\ret
	 & = &
	\sum_{\omega=1}^\Omega P_{\omega}
	\frac{
	\sum_{\ell\in I_{\omega}}\chi[j(\ell)=j]|\phi_{\ell}|^2
	}{
	\sum_{\ell\in I_{\omega}}|\phi_{\ell}|^2
	}.
	\label{phiPw}
\end{eqnarray}
Since we have $\sum_{\omega=1}^\Omega P_{\omega}=1$ and
$P_{\omega}\ge0$, (\ref{partB}) immediately implies (\ref{wholeB1})
with
\begin{equation}
	W_{j}=\sum_{\omega=1}^\Omega P_{\omega}W_{j}^{(\omega)}.
	\label{Wj}
\end{equation}
The normalization property (\ref{Wnorm}) is trivial if we note
(\ref{Wwj}) and (\ref{Wj}).
The bound (\ref{Wbound}) follows from (\ref{Wwj}) and (\ref{Wj}) if
we note
$E-\lambda\le U_{\tilde{\ell}(\omega)}\le E+\lambda$ for
$\omega=1,2,\ldots,\Omega$ and $\rho$ is nondecreasing.\qed

\bigskip
For any operator $A$ on the Hilbert space $\HiS$ of the subsystem
 (with the matrix elements
$(A)_{j,j'}=\bkt{\Psi_{j},A\Psi_{j'}}$), we define
\begin{equation}
	\bkt{A}_{W}=\sum_{j=1}^n(A)_{j,j}W_{j}.
	\label{AW}
\end{equation}
We show that this expectation value is almost equal to the desired
canonical expectation value
\begin{equation}
	\can{A}{\beta}
	=
	\frac{{\rm Tr}_{\rm S}[Ae^{-\beta\HS}]}
	{{\rm Tr}_{\rm S}[e^{-\beta\HS}]}
	=
	\frac{
	\sum_{j=1}^n(A)_{j,j}e^{-\beta\ep_{j}}
	}{
	\sum_{j=1}^ne^{-\beta\ep_{j}}
	}.
	\label{Acan}
\end{equation}
This is an elementary estimate, and can be
proved rather easily.
\begin{lemma}
\label{l:Wcan}
For any operator $A$ on $\HiS$, we have
\begin{equation}
	\abs{
	\bkt{A}_{W}-\can{A}{\beta}
	}
	\normi{A}\cbk{3\beta\gamma+\gamma(\ep_{n})^2},
	\label{Wcan}
\end{equation}
with $\beta=\beta(E)=d\log\rho(E)/dE$ and
$\gamma=d\beta(E+\lambda)/dE$.
\end{lemma}

\proof
Let $\tE_{j}=E-\ep_{j}$.
We shall prove upper and lower bounds for
$\rho(\tE_{j}\pm\lambda)/\rho(\tE_{1})$.
Since $\beta(E)=d\log\rho(E)/dE$, we have
\begin{equation}
	\frac{\rho(\tE)}{\rho(\tE_{1})}
	=
	\exp\sqbk{\int_{\tE_{1}}^{\tE} dE\,\beta(E)},
	\label{rho/rho}
\end{equation}
for any $\tE$.
By expanding $\beta(\tE)$ around $\tE_{1}$ and recalling that
$d\beta(E)/dE$ is increasing, we have
\begin{equation}
	0\le
	\int_{\tE_{1}}^{\tE}\,\beta(E)-(\tE-\tE_{1})\beta(\tE_{1})
	\le
	\frac{\gamma}{2}(\tE-\tE_{1})^2,
	\label{intdEbeta}
\end{equation}
where $\gamma=\beta'(\tE_{1}+\lambda)$.
Substituting these bounds into (\ref{rho/rho}), we have
\begin{eqnarray}
	\frac{\rho(\tE_{j}+\lambda)}{\rho(\tE_{1})}
	 & \le &
	e^{-\beta(\ep_{j}-\ep_{1})+\beta\lambda+
	(\gamma/2)(\ep_{j}-\ep_{1}-\lambda)^2}
	\ret
	 & \le &
	e^{-\beta(\ep_{j}-\ep_{1})+\beta\lambda+
	(\gamma/2)(\ep_{n}-\ep_{1})^2},
	\label{rr<}
\end{eqnarray}
and
\begin{equation}
	\frac{\rho(\tE_{j}-\lambda)}{\rho(\tE_{1})}
	\ge
	e^{-\beta(\ep_{j}-\ep_{1})-\beta\lambda},
	\label{rr>}
\end{equation}
with $\beta=\beta(\tE_{1})$ for any $j=1,2,\ldots,n$.
By substituting (\ref{rr<}), (\ref{rr>}) into (\ref{Wbound}), we
finally get
\begin{equation}
	e^{-2\beta\lambda-(\gamma/2)(\ep_{n}-\ep_{1})^2}
	\le
	\frac{W_{j}}{\widetilde{W}_{j}^{(\beta)}}
	\le
	e^{2\beta\lambda+(\gamma/2)(\ep_{n}-\ep_{1})^2},
	\label{W/W}
\end{equation}
where
$\widetilde{W}_{j}^{(\beta)}=e^{-\beta\ep_{j}}/
\sum_{j'=1}^ne^{-\beta\ep_{j'}}$
is the Boltzmann factor.

We can finish the proof of (\ref{Wcan}) by observing that
\begin{eqnarray}
	\abs{
	\bkt{A}_{W}-\can{A}{\beta}
	} & = &
	\abs{
	\sum_{j=1}^n(A)_{j,j}(W_{j}-\widetilde{W}_{j}^{(\beta)})
	}
	\ret
	 & \le &
	\cbk{
	\sum_{j=1}^n\abs{(A)_{j,j}}\widetilde{W}_{j}^{(\beta)}
	}
	\max_{j}\abs{
	\frac{W_{j}}{\widetilde{W}_{j}^{(\beta)}}-1
	}
	\ret
	 & \le &
	\normi{A}\cbk{3\beta\lambda+\gamma(\ep_{n}-\ep_{1})^2},
	\label{AW-Acan}
\end{eqnarray}
where the final line follows form (\ref{W/W}) if
$2\beta\lambda+(\gamma/2)(\ep_{n}-\ep_{1})^2\le0.7$, which we shall
assume.\qed

\bigskip
To complete the proof of the Lemma in the main paper, we need one
more estimate which will be proved in Section~\ref{s:decay}.
\begin{lemma}
\label{l:offD}
For any normalized $\phi_{\ell}$ satisfying (\ref{Sch}), we have
\begin{equation}
	\sum_{k=1}^N\overline{\phi_{(j,k)}}\phi_{(j',k)}
	\le
	{\rm const.}e^{-{\rm const.}L^{1/3}},
	\label{offD}
\end{equation}
for any $j\ne j'$.
\end{lemma}

Given all the above estimates, to prove the desired Lemma in the main
paper is straightforward.
With the expansion
\begin{equation}
	\Phi_{E}=\sum_{j=1}^n\sum_{k=1}^N\phi_{(j,k)}\,
	\Psi_{j}\otimes\Gamma_{k},
	\label{PhiEexpand}
\end{equation}
the desired quantity becomes
\begin{equation}
	\bkt{\Phi_{E},(A\otimes\oneB)\Phi_{E}}
	=
	\sum_{j,j'=1}^n(A)_{j,j'}
	\cbk{\sum_{k=1}^N\overline{\phi_{(j,k)}}\phi_{(j',k)}}.
	\label{EAE=App}
\end{equation}
Since the diagonal weigh $\sum_{k=1}^N|\phi_{(j,k)}|^2$ is controlled
by (\ref{wholeB1}), and the off diagonal weight
$\sum_{k=1}^N\overline{\phi_{(j,k)}}\phi_{(j',k)}$ for $j\ne j'$ by
(\ref{offD}), we immediately see that
\begin{equation}
	\abs{
	\bkt{\Phi_{E},(A\otimes\oneB)\Phi_{E}}
	-
	\bkt{A}_{W}
	}
	\le
	2c_{1}\normi{A}L^{-\eta}.
	\label{EAE-AW}
\end{equation}
By combining this with the systematic error estimate (\ref{Wcan}), we
get the desired (7) of the main paper.

\subsection{Regular intervals}
\label{s:reg}
We first recall special regular structure of the spectrum
$\{\Bk\}_{k=1,\ldots,N}$ of $\HB$.
In each interval $((r-1)\delta,r\delta)$ with $r=1,2,\ldots,R$, the
energy eigenvalues $\Bk$ are spaced with exactly equal spacing
$b_{r}=\delta(M_{r}L)^{-1}$.
This means that the whole index set is naturally decomposed into $R$
intervals as
$\{1,2,\ldots,N\}=\bigcup_{r=1}^RK_{R}$, so that for any $k\in K_{r}$,
we have $\Bk\in((r-1)\delta,r\delta)$.

Since $U_{(j,k)}=\epj+\Bk$, the structure of $\{U_{\ell}\}$ inherits
the above regularity of $\{\Bk\}$.
We say that an interval $J\subset\{1,2,\ldots,nN\}$ is {\em regular}
if for any $(j,k)\lra\ell\in J$, we have $k\in K_{r(j)}$ with
($J$-dependent) $r(1),r(2),\ldots,r(n)(=1,2,\ldots,R)$.
Thus, in the interval $J$, $U_{\ell}$ is constructed by superposing
$n$ shifted copies of $\{\Bk\}$ each of them having exactly equal
level spacings.

The whole range of $\ell$ can be decomposed into a disjoint union as
\begin{equation}
	\{1,2,\ldots,nN\}
	=
	\bigcup_{s=1}^{nR}J_{s},
	\label{decJ}
\end{equation}
where each $J_{s}$ is a maximal regular interval.
We note that each regular interval has length of $O(L)$.
We also remark that (\ref{decJ}) is not yet
the decomposition (\ref{dec}).

We want to determine the behavior of the index $\ell(j,k)$,
$U_{\ell}$, and $\alpha_{\ell}=(U_{\ell}-E)/\lambda$ in a fixed
regular interval $J$, which is one of $J_{1},\ldots,J_{nR}$.
For simplicity, we write
\begin{equation}
	\tb_{j}=b_{r(j)},
	\quad
	\tM_{j}=M_{r(j)},
	\label{bbMM}
\end{equation}
for $j=1,2,\ldots,n$.

Let $(j,k)\lra\ell\in J$.
Because of the regularity, we can write
\begin{equation}
	U_{(j,k)}=(k-\kappa_{j})\tb_{j}+u_{j},
	\label{Ujk}
\end{equation}
for some $\kappa_{j}$ and $u_{j}$ (which are again $J$-dependent).
The index $\ell$ is determined by ordering $U_{(j,k)}$ so that
$U_{\ell}\le U_{\ell+1}$.

To get an explicit formula for $\ell(j,k)$, we count the number of
$(j',k')\in J$ such that
\begin{equation}
	U_{(j',k')}\le U_{(j,k)}.
	\label{U<U}
\end{equation}
For a fixed $j'\ne j$, the number of $k'$ with (\ref{U<U}) is
\begin{equation}
	\sqbk{\frac{U_{(j,k)}-u_{j'}}{\tb_{j'}}},
	\label{[U-u]}
\end{equation}
where $[\cdots]$ is the Gauss symbol.
Summing up these contribution as well as that from the indexes
$(j,k')\in J$, we get
\begin{eqnarray}
	\ell(j,k) & = &
	\ell_{0}-1+(k-\kappa_{j})+1
	+\sum_{j'\ne j}\sqbk{\frac{U_{(j,k)}-u_{j'}}{\tb_{j'}}}
	\ret
	 & = &
	\ell_{0}+\sum_{j'=1}^n
	\sqbk{\frac{(k-\kappa_{j})\tb_{j}+u_{j}-u_{j'}}{\tb_{j'}}},
	\label{ljk1}
\end{eqnarray}
where we used (\ref{Ujk}), and $\ell_{0}$ is the smallest index in
$J$.
For the latter use, we substitute the relation
$\tb_{j}=\delta(\tM_{j}L)^{-1}$ into (\ref{ljk1}) to get
\begin{equation}
	\ell(j,k)=
	\ell_{0}'+\sum_{j'=1}^n
	\sqbk{\frac{\tM_{j'}}{\tM_{j}}k+\eta_{j.j'}},
	\label{ljk2}
\end{equation}
where $\ell_{0}'$ and $\eta_{j,j'}$ are constants (which may depend
on $L$).

From (\ref{ljk1}), we see that
\begin{equation}
	\tl(j,k)-(n-1)\le \ell(j,k)\le\tl(j,k),
	\label{l<l<l}
\end{equation}
with
\begin{equation}
	\tl(j,k)=
	\ell_{0}
	+\sum_{j'=1}^n
	\frac{(k-\kappa_{j})\tb_{j}+u_{j}-u_{j'}}{\tb_{j'}}.
	\label{ltjk}
\end{equation}
Let
\begin{equation}
	\bb=\rbk{\sum_{j=1}^n\frac{1}{\tb_{j}}}^{-1}.
	\label{bbar}
\end{equation}
Note that
\begin{equation}
	\bb=\rbk{\sum_{j=1}^n\frac{\tM_{j}L}{\delta}}^{-1}
	=\rbk{\frac{\delta}{\sum_{j=1}^n\tM_{j}}}L^{-1}
	=\frac{g\lambda}{L},
	\label{bbar2}
\end{equation}
where
\begin{equation}
	g=\frac{\delta}{\lambda\sum_{j=1}^n\tM_{j}}
	\label{gdef}
\end{equation}
is an $L$-independent
quantity.
Observe that
\begin{eqnarray}
	\bb\cbk{\tl(j,k)-\ell_{0}} & = &
	(k-\kappa_{j})\tb_{j}+u_{j}-\bu
	\ret
	 & = &
	U_{(j,k)}-\bu,
	\label{bll}
\end{eqnarray}
where
\begin{equation}
	\bu=\bb\sum_{j=1}^n\frac{u_{j}}{\tb_{j}}
	\label{ubar}
\end{equation}
is an energy near the bottom of the interval $J$.
From (\ref{bll}) and (\ref{l<l<l}), we get
\begin{equation}
	\bb\ell+\tU\le U_{\ell}\le\bb\ell+\tU+\bb(n-1),
	\label{<Ul<}
\end{equation}
for $\ell$ in the regular interval $J$, where $\tU=\bu-\bb\ell_{0}$.

We introduce a linearized $\alpha_{\ell}$ by
\begin{equation}
	\bal_{\ell}=\frac{\bb\ell+\tU-E}{\lambda}
	=g\frac{\ell}{L}+\frac{\tU-E}{\lambda},
	\label{alphabar}
\end{equation}
for $\ell\in J$.
Then from (\ref{<Ul<}), we find
\begin{equation}
	\abs{\alpha_{\ell}-\bal_{\ell}}
	\le
	\frac{\bb(n-1)}{\lambda}
	=\frac{g(n-1)}{L}.
	\label{a-abar}
\end{equation}

\subsection{Decomposition into intervals}
\label{s:int}
We describe precise definitions of the decomposition (\ref{dec}) of
the intervals.
The intervals $I_{1},\ldots,I_{\Omega}$ are properly ordered, and
covers the whole range $\{1,2,\ldots,nN\}$ without any overlaps.

We define the first turning point $\ellt$ as the minimum $\ell$ such
that $\alpha_{\ell}>-1$.
Then the first interval is defined as
\begin{equation}
	I_{1}=\cbk{1,2,\ldots,\ellt+[c_{2}L^{1/3}]}.
	\label{I1}
\end{equation}
Again $[\cdots]$ is the Gauss symbol.
Similarly we define
\begin{equation}
	I_{\Omega}=\cbk{\ellt'-[c_{2}L^{1/3}],\ldots,nN},
	\label{IOmega}
\end{equation}
where the second turning point $\ellt'$ is the maximum $\ell$ such
that $\alpha_{\ell}<1$.

The intervals $I_{2},\ldots,I_{\Gamma}$ all have the length
$|I_{\omega}|=[c_{3}L^{(1/3)-\eta}]$.
$\Gamma$ is determined as the minimum number such that
$\sum_{\omega=2}^\Gamma|I_{\omega}|\ge[c_{4}L^{1-\ep}]$.
The exponent $\ep>0$ will be determined later, but it must satisfy
\begin{equation}
	1-\ep>\frac{1}{3}-\eta,
	\label{cond0}
\end{equation}
because we must have
$\sum_{\omega'=2}^\Gamma|I_{\omega'}|\ge|I_{\omega}|$.
Similarly the intervals
$I_{\Omega-\Gamma+1},\ldots,I_{\Omega-1}$ have the length
$[c_{3}L^{(1/3)-\eta}]$.

The remaining intervals $I_{\Gamma+1},\ldots,I_{\Omega-\Gamma}$ are
defined to cover the (wide) remaining region
$\cbk{\ellt+[c_{2}L^{1/3}]+[c_{4}L^{1-\ep}],\ldots,
\ellt-[c_{2}L^{1/3}]-[c_{4}L^{1-\ep}]}$.
We require for $\omega=\Gamma+1,\ldots,\Omega-\Gamma$ that
$I_{\omega}$ has the length
\begin{equation}
	[c_{5}L^{1-2\theta}/2]\le|I_{\omega}|\le[c_{5}L^{1-2\theta}],
	\label{<Io<}
\end{equation}
where $\theta$ is an exponent to be determined later, and
$I_{\omega}$ is contained in a single regular interval defined in
Section~\ref{s:reg}.
These two conditions are easily satisfied since the lengths of the
regular intervals are of $O(L)$.

\subsection{Approximate solutions in the ``classically accessible
region''}
\label{s:discap}
We will construct approximate solutions of the \Sch\ equation
(\ref{Sch}) in the intervals $I_{2},\ldots,I_{\Omega-1}$, which are
within the ``classically accessible region''.

For simplicity, we denote by
$I=\cbk{\ell_{1},\ell_{1}+1,\cdots,\ell_{2}}$ one of the intervals
$I_{2},\ldots,I_{\Omega-1}$.
Since the interval $I$ is entirely contained in a single regular
interval $J$, we have corresponding $\bal_{\ell}$ defined by
(\ref{alphabar}), which is a linear approximation to $\alpha_{\ell}$.
We also write
\begin{equation}
	\alpha=\alpha_{\ell_{2}},\quad
	\beta=\alpha_{\ell_{1}}.
	\label{alphabeta}
\end{equation}

We start from an abstract theory for (rigorously)
evaluating the difference between an approximate solution
 and the true solution  of
(\ref{Sch}).
Let $\psi_{\ell}$ be an approximate solution of (\ref{Sch}) with
$\alpha_{\ell}$ replaced by its linearlization $\bal_{\ell}$ in the
sense that
\begin{equation}
	\psi_{\ell+1}+\psi_{\ell-1}+2\bal_{\ell}\psi_{\ell}
	=\delta_{\ell}
	\label{Sch2}
\end{equation}
holds for $\ell\in\bar{I}=\cbk{\ell_{1}+1,\ldots,\ell_{2}-1}$.
Here $\delta_{\ell}$ is an error term, which must be small.

For a solution $\phi_{\ell}$ of the original \Sch\ equation
(\ref{Sch}), we write the deviation from the approximate solution
as
\begin{equation}
	f_{\ell}=\phi_{\ell}-\psi_{\ell}.
	\label{fdef}
\end{equation}
From (\ref{Sch}) and (\ref{Sch2}), we find
\begin{equation}
	f_{\ell+1}+f_{\ell-1}+2\alpha_{\ell}f_{\ell}=\sigma_{\ell},
	\label{Schf}
\end{equation}
with
\begin{equation}
	\sigma_{\ell}=
	2(\bal_{\ell}-\alpha_{\ell})\psi_{\ell}-\delta_{\ell}.
	\label{sigma}
\end{equation}
As usual the equation (\ref{Schf}) can be written in a matrix form as
\begin{equation}
	\rbk{\matrix{f_{\ell+1}\cr f_{\ell}}}
	=
	\sT_{\ell}\rbk{\matrix{f_{\ell}\cr f_{\ell-1}}}
	+\rbk{\matrix{\sigma_{\ell}\cr0}},
	\label{Schfm}
\end{equation}
where the transfer matrix is
\begin{equation}
	\sT_{\ell}=\rbk{\matrix{
	-2\alpha_{\ell}&-1\cr1&0
	}}.
	\label{Tl}
\end{equation}

Considering the second order nature of the equations (\ref{Sch}) and
 (\ref{Sch2}), we can assume without loosing generality that
 $f_{\ell_{1}}=f_{\ell_{1}+1}=0$.
 Then by inserting (\ref{Schfm}), we find
 \begin{equation}
 	\rbk{\matrix{f_{\ell+1}\cr f_{\ell}}}
 	=
 	\sum_{k=\ell_{1}+1}^\ell
 	\sT_{\ell}\sT_{\ell-1}\cdots\sT_{k+1}
 	\rbk{\matrix{\sigma_{k}\cr0}}.
 	\label{frec}
 \end{equation}

 Since we have $|\alpha_{\ell}|<1$ in the ``classically accessible
 region'', the transfer matrix $\sT_{\ell}$ of (\ref{Tl}) has two
 eigenvalues $e_{\ell},\overline{e_{\ell}}$ with
 $|e_{\ell}|=|\overline{e_{\ell}}|=1$, where
\begin{equation}
	e_{\ell}=-\alpha_{\ell}+i\sqrt{1-{\alpha_{\ell}}^2}.
	\label{el}
\end{equation}
Thus there exists a regular matrix $\sM_{\ell}$ such that
\begin{equation}
	(\sM_{\ell})^{-1}\sT_{\ell}\sM_{\ell}=
	\rbk{\matrix{
	e_{\ell}&0\cr0&\overline{e_{\ell}}
	}}.
	\label{MTM}
\end{equation}
Let us define
\begin{equation}
	\sD_{\ell}=(\sM_{\ell})^{-1}\sM_{\ell-1}-
	\rbk{\matrix{1&0\cr0&1}},
	\label{Dl}
\end{equation}
which is expected to be small since $\sM_{\ell}$ and $\sM_{\ell-1}$
are very similar with each other.
Then from (\ref{MTM}), we have
\begin{eqnarray}
	 &  &
	 \sT_{\ell}\sT_{\ell-1}\cdots\sT_{k+1}
	\ret
	 &  & =
	\sM_{\ell}\eeD{\ell}(\sM_{\ell})^{-1}
	\sM_{\ell-1}\eeD{\ell-1}(\sM_{\ell-1})^{-1}
	\cdots
   \ret&&
   \cdots
	\sM_{k+1}\eeD{k+1}(\sM_{k+1})^{-1}
	\ret
	 &  & =
	\sM_{\ell}\eeD{\ell}(\one+\sD_{\ell})
	\eeD{\ell-1}(\one+\sD_{\ell-1})
	\cdots
   \ret&&
   \cdots
	(\one+\sD_{k+2})\eeD{k+1}(\sM_{k+1})^{-1}.
	\label{TTT}
\end{eqnarray}

For any $a,b=1,2$, we denote by $({\sf A})_{a,b}$ the $a,b$-component
of a $2\times2$ matrix $\sf A$.
We now assume that
\begin{equation}
	\abs{(\sD_{\ell})_{a,b}}\le d_{\ell},
	\label{D<}
\end{equation}
for any $a,b=1,2$.
Then (\ref{TTT}) implies
\begin{eqnarray}
	\abs{(\sT_{\ell}\sT_{\ell-1}\cdots\sT_{k+1})_{a,b}}
	& \le &
	\normi{\sM_{\ell}}\normi{(\sM_{k+1})^{-1}}
	\prod_{j=k+1}^\ell(1+2d_{j})
	\ret
	 & \le &
	\normi{\sM_{\ell}}\normi{(\sM_{k+1})^{-1}}
	\exp\sqbk{2\sum_{j=k+1}^\ell d_{j}} .
	\label{TTT<}
\end{eqnarray}
To make the estimate (\ref{TTT<}) more concrete, we fix precise form
of $\sM_{\ell}$ as
\begin{equation}
	\sM_{\ell}=\frac{1}{\sqrt{2}(1-{\alpha_{\ell}}^2)^{1/4}}
	\rbk{\matrix{e_{\ell}&\overline{e_{\ell}}\cr1&1}},
	\label{Ml}
\end{equation}
whose inverse is
\begin{equation}
	(\sM_{\ell})^{-1}
	=\frac{1}{i\sqrt{2}(1-{\alpha_{\ell}}^2)^{1/4}}
	\rbk{\matrix{1&-\overline{e_{\ell}}\cr-1&e_{\ell}}}.
	\label{Mli}
\end{equation}

Since $\sD_{\ell}$ defined by (\ref{Dl}) is vanishing if
$\alpha_{\ell}=\alpha_{\ell-1}$, we can evaluate $\sD_{\ell}$ by
expanding it in $(\alpha_{\ell}-\alpha_{\ell-1})$.
Then we find that essential contribution comes from the first order in
the expansion, and the desired bound (\ref{D<}) is satisfied with
\begin{equation}
	d_{\ell}=\frac{\alpha_{\ell}-\alpha_{\ell-1}}
	{1-{\alpha_{\ell}}^2},
	\label{dl}
\end{equation}
provided that
\begin{equation}
	1-{\alpha_{\ell}}^2\ge\frac{2}{L},
	\label{1-a2}
\end{equation}
which is automatically satisfied in the present interval.
Substituting (\ref{dl}) into (\ref{TTT<}), we find
\begin{equation}
	\abs{(\sT_{\ell}\sT_{\ell-1}\cdots\sT_{k+1})_{a,b}}
	\le
	\normi{\sM_{\ell}}\normi{(\sM_{k+1})^{-1}}
	\exp\sqbk{\frac{2(\alpha_{\ell}-\alpha_{k+1})}{1-\gamma^2}},
	\label{TTT<2}
\end{equation}
for any $a,b=1,2$,
where $\gamma$ is one of $\alpha_{\ell}$ with $\ell\in I$ which gives
the smallest $1-{\alpha_{\ell}}^2$.

At this stage, we impose a condition on $I$ that
\begin{equation}
	\frac{\alpha-\beta}{1-\gamma^2}\le c_{6},
	\label{a-b}
\end{equation}
where $\alpha,\beta$ are defined by (\ref{alphabeta}).
Then (\ref{TTT<2}) simplifies as
\begin{eqnarray}
	\abs{(\sT_{\ell}\sT_{\ell-1}\cdots\sT_{k+1})_{a,b}}
	& \le &
	\normi{\sM_{\ell}}\normi{(\sM_{k+1})^{-1}}e^{c_{6}}
	\ret
	 & \le &
	c_{7}(1-{\alpha_{\ell}}^2)^{-1/4}(1-{\alpha_{k}}^2)^{-1/4},
	\label{TTT<3}
\end{eqnarray}
where we used (\ref{Ml}) and (\ref{Mli}).
Substituting (\ref{TTT<3}) to (\ref{frec}), we get a useful bound
\begin{equation}
	|f_{\ell+1}|
	\le
	\frac{c_{7}}{(1-{\alpha_{\ell}}^2)^{1/4}(1-\gamma^2)^{1/4}}
	\sum_{k=\ell_{1}+1}^\ell|\sigma_{k}|.
	\label{f<}
\end{equation}
This completes the general theory.

We now explicitly construct an approximate solution $\psi_{\ell}$
which appears in (\ref{Sch2}).
We first define $\zeta_{\ell}$ by
\begin{equation}
	\cos\zeta_{\ell}=-\bal_{\ell}.
	\label{zeta}
\end{equation}
Since $\bal_{\ell}$ is increasing in $\ell$ and satisfies
$|\bal_{\ell}|<1$, we see that $\zeta_{\ell}$ is increasing and
satisfies $0<\zeta_{\ell}<\pi$.
Let us define
\begin{equation}
	Z_{\ell}=
	\rbk{\sum_{\ell'=\ell_{1}}^{\ell-1}\zeta_{\ell'}}
	+\frac{\zeta_{\ell}}{2}.
	\label{Zl}
\end{equation}
Then we can write down our approximate solution as
\begin{equation}
	\psi_{\ell}=\frac{\cos(Z_{\ell}+\xi)}{\sqrt{\sin\zeta_{\ell}}},
	\label{psi}
\end{equation}
where $\xi$ is an arbitrary constant.
The form (\ref{psi}) is (heuristically) obtained by following the
standard idea of quasi-classical analysis.
We learned that
this approximate solution was written down long time ago by
Bethe\footnote{
H. Bethe, Phys. Rev. {\bf 54}, 955 (1938).
}.

We now need an estimate for the error term $\delta_{\ell}$ which
appears in (\ref{Sch2}).
To do this, we substitute the concrete form (\ref{psi}) into the
left-hand side of
(\ref{Sch2}), and expand the resulting quantity in a power series
of
$(\bal_{\ell+1}-\bal_{\ell})=(\bal_{\ell}-\bal_{\ell-1})
=\bb/\lambda(=g/L)$.
(See (\ref{alphabar}.)
We find that the first order terms cancel out, and the essential
contribution comes from the second order.
We skip the tedious but straightforward calculation, and describe
only the final result, which is
\begin{eqnarray}
	|\delta_{\ell}|
	& \le &
	\frac{1}{2(1-(\bal_{\ell})^2)^{3/4}}
	\frac{\alpha_{\ell}}{(1-(\bal_{\ell})^2)^{3/2}}
	\rbk{\frac{\bb}{\lambda}}^2
	\ret
	 &  &
	 +\frac{2}{(1-(\bal_{\ell})^2)^{5/4}}
	 \rbk{\frac{\bb}{\lambda(1-(\bal_{\ell})^2)^{1/2}}}^2
	\ret
	 & \le &
	\frac{3}{(1-(\bal_{\ell})^2)^{9/4}}\rbk{\frac{\bb}{\lambda}}^2.
	\label{delta<}
\end{eqnarray}
Here we used the fact that $\alpha_{\ell}\simeq\bal_{\ell}$ in the
sense of (\ref{a-abar}).

By recalling the definition (\ref{sigma}) of $\sigma_{\ell}$, and
using (\ref{delta<}) and (\ref{a-abar}), we find
\begin{equation}
	|\sigma_{\ell}|\le
	\frac{2n}{(1-{\alpha_{\ell}}^2)^{1/4}}
	\rbk{\frac{\bb}{\lambda}}
	+\frac{3}{(1-{\alpha_{\ell}}^2)^{9/4}}
	\rbk{\frac{\bb}{\lambda}}^2,
	\label{sigma<}
\end{equation}
where we noted that $\sin\zeta_{\ell}=\sqrt{1-{\alpha_{\ell}}^2}$.
Since
\begin{equation}
	\sum_{k=\ell_{1}+1}^\ell\frac{\bb}{\lambda}
	=\bal_{\ell}-\bal_{\ell_{1}}
	\simeq\alpha_{\ell}-\alpha_{\ell_{1}},
	\label{sumne}
\end{equation}
we find
\begin{equation}
	\sum_{k=\ell_{1}+1}^\ell|\sigma_{k}|
	\le
	\frac{2n(\alpha-\beta)}{(1-\gamma^2)^{1/4}}
	+\frac{3(\alpha-\beta)}{(1-\gamma^2)^{9/4}}
	\rbk{\frac{\bb}{\lambda}}.
	\label{sumsigma<}
\end{equation}

Going back to (\ref{f<}), we finally see that the relative error is
bounded as
\begin{eqnarray}
	\abs{f_{\ell}\sqrt{\sin\zeta_{\ell}}}
	& \le &
	\frac{c_{7}}{(1-\gamma^2)^{1/4}}
	\sum_{k=\ell_{1}+1}^\ell|\sigma_{k}|
	\ret
	 & \le &
	\frac{2nc_{7}(\alpha-\beta)}{(1-\gamma^2)^{1/2}}
	+\frac{3c_{7}(\alpha-\beta)}{(1-\gamma^2)^{5/2}}
	\rbk{\frac{\bb}{\lambda}}.
	\label{reler}
\end{eqnarray}
For the latter uses, we want to get a bound of the form
\begin{equation}
	\abs{f_{\ell}\sqrt{\sin\zeta_{\ell}}}
	\le
	c_{8}L^{-\eta},
	\label{reler2}
\end{equation}
with a constant $c_{8}$ which does not depend on $L$ and the specific
interval.

We shall write down conditions that the exponents $\eta$, $\ep$ and
$\theta$ should satisfy to get (\ref{reler2}).
We start from the case where $I$ is one of the second type intervals,
i.e., $I_{\omega}$ with $\omega=2,\ldots,\Gamma$ or
$\omega=\Omega-\Gamma+1,\ldots,\Omega-1$.
These are the intervals which are relatively close to the turning
points.
For such an interval, we have
$1-\gamma^2\ge c_{2}L^{1/3}(\bb/\lambda)=c_{2}gL^{-2/3}$.
(See (\ref{gdef}) for the definition of $g$, which is an
$L$-independent quantity.)
We also find
$(\alpha-\beta)\le c_{3}L^{(1/3)-\eta}(\bb/\lambda)
=c_{3}gL^{-(2/3)-\eta}$.
It is easy to check that the condition (\ref{a-b}) is satisfied.
By substituting these bound into (\ref{reler}), we see that
\begin{equation}
	\abs{f_{\ell}\sqrt{\sin\zeta_{\ell}}}
	\le
	2nc_{7}c_{3}\sqrt{\frac{g}{c_{2}}}L^{-(1/3)-\eta}
	+\frac{3c_{7}c_{3}}{\sqrt{{c_{2}}^5g}}L^{-\eta}.
	\label{reler3}
\end{equation}
Thus we find that the desired bound (\ref{reler2}) is indeed
satisfied.

We then consider the case where $I$ is one of the third type
intervals, i.e., $I_{\omega}$ with
$\omega=\Gamma+1,\ldots,\Omega-\Gamma$.
Here we have
$(1-\gamma^2)\ge c_{4}L^{1-\ep}(\bb/\lambda)=c_{4}gL^{-\ep}$, and
$(\alpha-\beta)\le c_{5}L^{1-2\theta}(\bb/\lambda)
=c_{5}gL^{-2\theta}$.
Therefore the condition (\ref{a-b}) is satisfied if
\begin{equation}
	-2\theta+\ep\le0.
	\label{cond0.5}
\end{equation}
Substituting these bounds into (\ref{reler}), we get
\begin{eqnarray}
	\abs{f_{\ell}\sqrt{\sin\zeta_{\ell}}}
	& \le &
	2nc_{7}c_{5}\sqrt{\frac{g}{c_{4}}}L^{-2\theta+(\ep/2)}
	+\frac{3c_{7}c_{5}}{\sqrt{{c_{4}}^5g}}L^{-1-2\theta+(5\ep/2)}.
	\ret
	 & = &
	\rbk{2nc_{7}c_{5}\sqrt{\frac{g}{c_{4}}}
	+\frac{3c_{7}c_{5}}{\sqrt{{c_{4}}^5g}}L^{2\ep-1}}
	L^{-2\theta+(\ep/2)}.
	\label{reler4}
\end{eqnarray}
Therefore, we have the desired bound (\ref{reler}) provided that
\begin{equation}
	\ep\le\frac{1}{2},
	\label{cond1}
\end{equation}
and
\begin{equation}
	-2\theta+\frac{\ep}{2}\le-\eta.
	\label{cond2}
\end{equation}
Later in Section~\ref{s:exp}, we determine all the exponents appear
in the proof.
As these exponents, we will set $\eta=\theta=1/6$ and $\ep=1/3$, which
satisfy (\ref{cond1}) and (\ref{cond2}).

To summarize, we have proved
\begin{lemma}
\label{l:discap}
Let $\phi_{\ell}$ be a real solution of (\ref{Sch}).
Let $I$ be one of the intervals $I_{\omega}$ with
$\omega=2,\ldots,\Omega-1$.
There are real constants $A$ and $\xi$, and we have
\begin{equation}
	\abs{
	\phi_{\ell}-A\frac{\cos(Z_{\ell}+\xi)}{\sqrt{\sin\zeta_{\ell}}}
	}
	\le
	c_{8}\frac{A}{\sqrt{\sin\zeta_{\ell}}}L^{-\eta},
	\label{discap}
\end{equation}
for $\ell\in I$.
\end{lemma}

\subsection{Approximate solutions near the ``turning points''}
\label{s:contap}
We will construct approximate solutions of the \Sch\ equation
(\ref{Sch}) in the ``classically accessible parts'' of the intervals
$I_{1}$ and $I_{\Omega}$.
The remaining ``classically inaccessible parts'' will be discussed in
Section~\ref{s:decay}.
Near the turning points, the wave length of the oscillation of
$\phi_{\ell}$ becomes long and the quasi-classical approximation
becomes useless.
Instead we try to approximate $\phi_{\ell}$ by the solution of a
rescaled
continuous \Sch\ equation that corresponds to (\ref{Sch}).

Let us discuss the interval $I_{1}$ which contains the first turning
point $\ellt$ at which $\alpha_{\ellt}\simeq1$.
The treatment of $I_{\Omega}$ is essentially the same.
We want to construct approximate solution of $\phi_{\ell}$ for $\ell$
in the interval
$\tilde{I}_{1}=\cbk{\ellt,\ellt+1,\ldots,\ellt+[c_{2}L^{1/3}]}
\subset I_{1}$.
let us write
\begin{equation}
	\beta_{\ell}=2(\alpha_{\ell}+1),
	\label{betal}
\end{equation}
and rewrite the \Sch\ equation (\ref{Sch}) as
\begin{equation}
	\phi_{\ell-1}-2\phi_{\ell}+\phi_{\ell+1}+\beta_{\ell}\phi_{\ell}
	=0.
	\label{Sch3}
\end{equation}
From (\ref{alphabar}) and (\ref{a-abar}), we find that
\begin{equation}
	\abs{\beta_{\ell}-\rbk{\frac{\bb}{\lambda}}(\ell-\ellt)}
	\le
	\frac{2(n-1)\bb}{\lambda}
	=2(n-1)gL^{-1}.
	\label{betaap}
\end{equation}

We want to approximate $\phi_{\ell}$ by a continuous function
$\psi(x)$ with $x=L^{-1/3}(\ell-\ellt)$.
With this correspondence in mind, we divide (\ref{Sch3}) by
$(L^{-1/3})^2$ to (roughly) get
\begin{equation}
	\psi''(x)+L^{-2/3}\beta_{\ell}\psi(x)\simeq0.
	\label{Sch3.5}
\end{equation}
We then note that
\begin{equation}
	L^{-2/3}\beta_{\ell}\simeq L^{-2/3}\frac{g}{L}(\ell-\ellt)
	=gx.
	\label{Lb=gx}
\end{equation}
So we are motivated to consider the continuous \Sch\ equation
\begin{equation}
	\psi''(x)+gx\,\psi(x)=0.
	\label{Sch4}
\end{equation}

Let $\psi(x)$ be the solution of (\ref{Sch4}) in the region $x\ge0$.
By an explicit calculation, one finds that two independent complex
solutions of (\ref{Sch4}) are given by
\begin{equation}
	\psi(x)=\sqrt{x}\,h[(2/3)\sqrt{g}x^{3/2}],
	\label{psih}
\end{equation}
and $\overline{\psi(x)}$, where
\begin{equation}
	h(z)=H_{1/3}^{(1)}(z)
	=J_{1/3}(z)+iY_{1/3}(z)
	\label{Hank}
\end{equation}
is the Hankel function (or Bessel's function of the third kind).
From the asymptotic behavior of the Hankel function, we find
\begin{equation}
	\psi(x)\approx x^{-1/4}
	\exp\sqbk{i\cbk{(2/3)\sqrt{g}x^{3/2}-(5\pi/12)}},
	\label{psiasymp}
\end{equation}
for $x\gg1$.

From now on, we use the index $m=\ell-\ellt$ for convenience.
To control the approximation rigorously, we first Taylor expand
$\psi(L^{-1/3}(m\pm1))$ to get
\begin{equation}
	\psi(L^{-1/3}(m+1))-2\psi(L^{-1/3}m)+\psi(L^{-1/3}(m-1))
	=
	L^{-2/3}\psi''(L^{-1/3}m)+L^{-1}\nu_{m},
	\label{psipsipsi}
\end{equation}
where
\begin{equation}
	\nu_{m}=
	\frac{1}{6}\cbk{\psi'''(x')+\psi'''(x'')}
	\simeq
	\frac{1}{3}\psi'''(L^{-1/3}m),
	\label{nu}
\end{equation}
with
$L^{-1/3}(m+1)\le x'\le L^{-1/3}m\le x''\le L^{-1/3}(m-1)$.
By using (\ref{Sch4}) and (\ref{psipsipsi}), we find
\begin{equation}
	\psi(L^{-1/3}(m+1))-2\psi(L^{-1/3}m)+\psi(L^{-1/3}(m-1))
	=
	-\frac{\bb}{\lambda}m\psi(L^{-1/3}m)+L^{-1}\nu_{m},
	\label{psieq}
\end{equation}
where we noted that
$L^{-2/3}gL^{-1/3}m=(\bb/\lambda)m$.
We divide the equation (\ref{psieq}) by $\psi(L^{-1/3}(m+1))$ to get
\begin{eqnarray}
	&&
	1-2\frac{\psi(L^{-1/3}m)}{\psi(L^{-1/3}(m+1))}+
	\frac{\psi(L^{-1/3}(m-1))}{\psi(L^{-1/3}(m+1))}
	\ret&&
	=
	-\frac{\bb}{\lambda}m
	\frac{\psi(L^{-1/3}m)}{\psi(L^{-1/3}(m+1))}
	+L^{-1}\frac{\nu_{m}}{\psi(L^{-1/3}(m+1))}.
	\label{psieq2}
\end{eqnarray}

We now let $\psi(x)$ be the specific complex solution (\ref{psih}),
and try to control a complex solution of (\ref{Sch}) such that
$\phi_{\ellt+m}\simeq\psi(L^{-1/3}m)$.
Information about the desired real solution can be read off easily,
as we do at the end of the present section.

Let us introduce a complex quantity $F_{m}$ by
\begin{equation}
	\phi_{\ellt+m}=F_{m}\psi(L^{-1/3}m).
	\label{Fdef}
\end{equation}
Then the \Sch\ equation (\ref{Sch}) becomes
\begin{eqnarray}
	&&
	F_{m+1}\psi(L^{-1/3}(m+1))-2F_{m}\psi(L^{-1/3}m)
	+F_{m-1}\psi(L^{-1/3}(m-1))
	\ret&&
	=
	-\beta_{\ellt+m}F_{m}\psi(L^{-1/3}m).
	\label{Fpsi1}
\end{eqnarray}
We divide this by $F_{m}\psi(L^{-1/3}(m+1))$to get
\begin{eqnarray}
	&&
	\frac{F_{m+1}}{F_{m}}
	-2\frac{\psi(L^{-1/3}m)}{\psi(L^{-1/3}(m+1))}
	+\frac{F_{m-1}}{F_{m}}
	\frac{\psi(L^{-1/3}(m-1))}{\psi(L^{-1/3}(m+1))}
	\ret&&
	=
	-\beta_{\ellt+m}
	\frac{\psi(L^{-1/3}m)}{\psi(L^{-1/3}(m+1))}.
	\label{Fpsi2}
\end{eqnarray}

From (\ref{psieq2}) and (\ref{Fpsi2}), we get the following recursion
equation for $F_{m}$.
\begin{eqnarray}
	\frac{F_{m+1}}{F_{m}}-1
	& = &
	-\frac{\psi(L^{-1/3}(m-1))}{\psi(L^{-1/3}(m+1))}
	\rbk{\frac{F_{m-1}}{F_{m}}-1}
	\ret
	 &  &
	-\rbk{\beta_{\ellt+m}-\frac{\bb}{\lambda}m}
	\frac{\psi(L^{-1/3}m)}{\psi(L^{-1/3}(m+1))}
	-\frac{\nu_{m}}{\psi(L^{-1/3}(m+1))}.
	\label{FF}
\end{eqnarray}
We now use the asymptotic behavior (\ref{psiasymp}) of $\psi(x)$ to
see that
\begin{eqnarray}
	\abs{\frac{\psi(L^{-1/3}(m-1))}{\psi(L^{-1/3}(m+1))}}
	& \le &
	1+{\rm const.}\abs{\frac{\psi'(x)}{\psi(x)}}L^{-1/3}
	\ret
	 & \le &
	1+(c_{9}+c_{10}x^{1/2})L^{-1/3}
	\ret
	 & = &
	1+c_{9}L^{-1/3}+c_{10}L^{-1/2}m^{1/2},
	\label{psi/psi1}
\end{eqnarray}
\begin{equation}
	\abs{\frac{\psi(L^{-1/3}m)}{\psi(L^{-1/3}(m+1))}}
	\le
	1+c_{9}L^{-1/3}+c_{10}L^{-1/2}m^{1/2},
	\label{psi/psi2}
\end{equation}
and
\begin{eqnarray}
	\abs{\frac{\nu_{m}}{\psi(L^{-1/3}(m+1))}}
	& \le &
	{\rm const.}
	\abs{\frac{\psi'''(x)}{\psi(x)}}
	\ret
	 & \le &
	c_{11}+c_{12}x^{3/2}
	\ret
	 & = &
	c_{11}+c_{12}L^{-1/2}m^{3/2}.
	\label{psi/psi3}
\end{eqnarray}

Let us define
\begin{equation}
	G_{m}=\frac{F_{m}}{F_{m-1}}-1.
	\label{Gdef}
\end{equation}
By using the estimates (\ref{psi/psi1}), (\ref{psi/psi2}),
(\ref{psi/psi3}) as well as (\ref{betaap}), the
recursion relation (\ref{FF}) reduces to
\begin{eqnarray}
	\abs{G_{m+1}}
	& \le &
	\rbk{|G_{m}|+|G_{m}|^2}
	(1+c_{9}L^{-1/3}+c_{10}L^{-1/2}m^{1/2})
	\ret
	 &  &
	+c_{13}L^{-1}
	(1+c_{9}L^{-1/3}+c_{10}L^{-1/2}m^{1/2})
	\ret
	 &  &
	+c_{11}L^{-1}+c_{12}L^{-3/2}m^{3/2},
	\label{Grec}
\end{eqnarray}
where we used
\begin{equation}
	\abs{\frac{F_{m-1}}{F_{m}}-1}
	=
	\abs{(G_{m}+1)^{-1}-1}
	\le
	|G_{m}|+|G_{m}|^2.
	\label{FFGG}
\end{equation}

We now assume that $G_{1}=0$.
Then we can estimate $|G_{m}|$ by repeatedly using (\ref{Grec}).
Let us assume $|G_{m}|\le\bG$ holds for any $m$ with
$1\le m\le c_{2}L^{1/3}$ where the constant $\bG$ will be determined
later.
Then (\ref{Grec}) implies for any $m$ with $1\le m\le c_{2}L^{1/3}$
that
\begin{eqnarray}
	|G_{m}|
	& \le &
	\sum_{j=1}^m\cbk{|G_{j}|-|G_{j-1}|}
	\ret
	 & \le &
	\bG(c_{2}c_{9}+{c_{2}}^{3/2}c_{10})
	\ret
	 &  &
	+\bG^2(c_{2}L^{1/3}+c_{2}c_{9}+{c_{2}}^{3/2}c_{10})
	\ret
	 &  &
	+c_{2}c_{13}(L^{-2/3}+c_{2}c_{9}L^{-1}
	+{c_{2}}^{3/2}c_{10}L^{-1})
	\ret
	 &  &
	+c_{2}c_{11}L^{-2/3}+{c_{2}}^{5/2}c_{12}L^{-2/3}.
	\label{Grec2}
\end{eqnarray}
We now set
\begin{equation}
	\bG=c_{14}L^{-2/3},
	\label{Gbar}
\end{equation}
and substitute this relation into (\ref{Grec2}).
We then find that, for sufficiently small (but $L$-independent)
$c_{2}$, (\ref{Grec2}) reproduces $|G_{m}|\le\bG$ with the same $\bG$.
This proves the upper bound
\begin{equation}
	|G_{m}|\le c_{14}L^{-2/3},
	\label{G<}
\end{equation}
for $m$ with $1\le m\le c_{2}L^{1/3}$.

Assuming $F_{0}=1$, we finally get
\begin{eqnarray}
	|F_{m}-1|
	& = &
	\abs{\rbk{\prod_{j=1}^m\frac{F_{j}}{F_{j-1}}}-1}
	\ret
	 & = &
	\abs{\cbk{\prod_{j=1}^m(1+G_{j})}-1}
	\ret
	 & \le &
	\exp[mc_{14}L^{-2/3}]-1
	\ret
	 & \le &
	c_{15}L^{-1/3},
	\label{F-1<}
\end{eqnarray}
for $m$ with $1\le m\le c_{2}L^{1/3}$.
Recalling (\ref{Fdef}), we have established the desired relation
$\phi_{\ellt+m}\simeq\psi(L^{-1/3}m)$.

To control the desired real solution of (\ref{Sch}), we only have to
sum up the complex solution and its complex conjugate with
appropriate complex weights.
This proves
\begin{lemma}
\label{l:contap}
Let $\phi_{\ell}$ be a real solution of (\ref{Sch}).
Then there is a complex constant $A$, and we have
\begin{eqnarray}
	&&
	\abs{
	\phi_{\ell}-
	\cbk{A\psi(L^{-1/3}(\ell-\ellt))
	+\overline{A\psi(L^{-1/3}(\ell-\ellt))}}
	}
	\ret&&
	\le
	2c_{15}|A|\abs{\psi(L^{-1/3}(\ell-\ellt))}L^{-1/3},
	\label{contap}
\end{eqnarray}
for $\ell\in\tilde{I}_{1}=\cbk{\ellt,\ldots,\ellt+[c_{2}L^{1/3}]}$,
where $\psi(x)$ is explicitly given by (\ref{psih}).
\end{lemma}

We still have to treat the solution in the ``classically accessible
part'' of the interval $I_{\Omega}$, i.e.,
$\tilde{I}_{\Omega}=\cbk{\ellt'-[c_{2}L^{1/3}],\ldots,\ellt'}$.
Since the analysis is exactly the same as that for $\tilde{I}_{1}$, we
only present the final result.
\begin{lemma}
\label{l:contap2}
Let $\phi_{\ell}$ be a real solution of (\ref{Sch}).
Then there is a complex constant $B$, and we have
\begin{eqnarray}
	&&
	\abs{
	\phi_{\ell}-
	(-1)^\ell\cbk{B\psi(L^{-1/3}(\ellt'-\ell))
	+\overline{B\psi(L^{-1/3}(\ellt'-\ell))}}
	}
	\ret&&
	\le
	2c_{15}|B|\abs{\psi(L^{-1/3}(\ellt'-\ell))}L^{-1/3},
	\label{contap2}
\end{eqnarray}
for $\ell\in\tilde{I}_{\Omega}$,
where $\psi(x)$ is explicitly given by (\ref{psih}).
\end{lemma}

\subsection{Decay of the solution in the
``classically inaccessible regions''}
\label{s:decay}
We now study the solution in the ``classically inaccessible region'',
which is characterized by $|\alpha_{\ell}|>1$.
Since the \Sch\ equation (\ref{Sch}) implies
\begin{equation}
	\phi_{\ell}=
	-\frac{\phi_{\ell-1}+\phi_{\ell-1}}{2\alpha_{\ell}},
	\label{Schphi}
\end{equation}
we get a convexity inequality
\begin{equation}
	|\phi_{\ell}|
	\le
	\frac{|\phi_{\ell-1}|+|\phi_{\ell-1}|}{2|\alpha_{\ell}|}
	<
	\frac{|\phi_{\ell-1}|+|\phi_{\ell-1}|}{2}.
	\label{convex}
\end{equation}
This means that $|\phi_{\ell}|$ cannot take a local maximum.

Let us focus on the region
$I_{\Omega}'=\cbk{\ellt'+1,\ldots,nN}\subset I_{\Omega}$.
Then (\ref{convex}) means that $|\phi_{\ell}|$ is decreasing because
of the boundary condition $\phi_{nN+1}=0$.
Then (\ref{convex}) further implies
\begin{equation}
	|\phi_{\ell}|
	\le
	\frac{|\phi_{\ell-1}|}{|\alpha_{\ell}|},
	\label{mono}
\end{equation}
and hence
\begin{equation}
	|\phi_{\ell}|
	\le
	|\phi_{\ellt'}|
	\prod_{\ell'=\ellt'}^{\ell}|\alpha_{\ell'}|^{-1}.
	\label{phi<a}
\end{equation}

We note that (\ref{a-abar}) implies
\begin{equation}
	|\alpha_{\ell'}|^{-1}
	\le
	\rbk{
	1+\frac{g}{L}(\ell'-\ellt')-\frac{ng}{L}
	}^{-1}
	\le
	\exp\sqbk{-c_{16}(g/L)(\ell'-\ellt'-n)},
	\label{ainv<}
\end{equation}
for $\ell'-\ellt'\le c_{17}L^{1-\mu}$ and any $\mu>0$.
Recall that $g$ is $L$-independent as in (\ref{gdef}).
Substituting (\ref{ainv<}) into (\ref{phi<a}), we get
\begin{equation}
	|\phi_{\ell}|\le|\phi_{\ellt'}|c_{18}
	\exp\sqbk{-c_{16}\frac{g}{L}(\ell'-\ellt')^{2}},
	\label{phi<philt}
\end{equation}
for $\ell'-\ellt'\le c_{17}L^{1-\mu}$.
This means that $|\phi_{\ell}|$ decays very rapidly in the
``classically inaccessible region''.
For $\ell\ge c_{17}L^{1-\mu}$, we have
\begin{equation}
	\frac{|\phi_{\ell}|}{|\phi_{\ellt'}|}
	\le
	c_{18}\exp\sqbk{
	-c_{16}{c_{17}}^{2}\frac{g}{2}L^{1-2\mu}
	}.
	\label{thedecay}
\end{equation}
In the most applications
(see (\ref{phik}) and Lemma~\ref{l:offD}),
we set $\mu=1/3$ in (\ref{thedecay}).

\subsection{Boltzmann factor from the interior of the
``classically accessible region''}
\label{s:res}
We are now ready to prove the most important Lemma~\ref{l:part}.
We first treat the intervals $I_{\omega}$ with
$\omega=\Gamma+1,\ldots,\Omega-\Gamma$.
These intervals are located in the interior of the ``classically
accessible region'', where the wave length of $\phi_{\ell}$ is
relatively short.
In such situations, we must face possible ``resonances'' between the
oscillation of $|\phi_{\ell}|^{2}$ and the quasi periodic behavior
of $\ell(j,k)$ (with $j$ fixed and $k$ varied), which (locally)
destroys the desired ``equal weighted'' behavior.
We will prove that such resonances are located in short intervals and
do not have significant contributions.

Let $I$ be one of $I_{\omega}$ with
$\omega=\Gamma+1,\ldots,\Omega-\Gamma$.
Our final goal is to evaluate the quantity
$S_{j}/(\sum_{j'=1}^{n}S_{j'})$ for $j=1,2,\ldots,n$,
where
\begin{equation}
	S_{j}=\sum_{\ell\in I}\chi[j(\ell)=j]|\phi_{\ell}|^{2}.
	\label{Sj}
\end{equation}
Because of the definition of the interval, we have
\begin{equation}
	1-{\alpha_{\ell}}^{2}\ge c_{4}gL^{-\ep},
	\label{1-a>}
\end{equation}
for any $\ell\in I$.
(See the end of Section~\ref{s:discap}.)
We also recall that the length of $I$ satisfies
\begin{equation}
	[c_{5}L^{1-2\theta}/2]\le|I|\le[c_{5}L^{1-2\theta}].
	\label{|I|}
\end{equation}

To evaluate the sum $S_{j}$ explicitly, we further decompose $I$ into
subintervals as
\begin{equation}
	I=\bigcup_{q=1}^{Q}\hI_{q},
	\label{I=Iq}
\end{equation}
with each $\hI_{q}$ having the length $|\hI_{q}|=[c_{19}L^{\nu}]$,
where $\nu>0$ is another exponent to be determined later (to be
$1/3$).
By $\hell_{q}$ we denote the smallest element in $\hI_{q}$.
Consequently, we write
\begin{equation}
	S_{j}=\sum_{q=1}^{Q}S_{j}^{(q)},
	\label{Sj2}
\end{equation}
with
\begin{equation}
	S_{j}^{(q)}
	=
	\sum_{\ell\in\hI_{q}}
	\chi[j(\ell)=j]|\phi_{\ell}|^{2}.
	\label{Sjq}
\end{equation}

Let $\ell\in\hI_{q}$.
We note that
\begin{eqnarray}
	\abs{\frac{\sin\zeta_{\ell}}{\sin\zeta_{\hell_{q}}}-1}
	& = &
	\abs{
	\frac{\sqrt{1-{\alpha_{\ell}}^{2}}}
	{\sqrt{1-{\alpha_{\hell_{q}}}^{2}}}
	-1
	}
	\ret
	 & \le &
	\rbk{\max_{\ell\in\hI_{q}}(1-{\alpha_{\ell}}^{2})^{-1}}
	(\alpha_{\ell}-\alpha_{\hell_{q}})
	\ret
	 & \le &
	\frac{c_{19}}{c_{4}}L^{\nu+\ep-1}
	\ret
	 & \le &
	\frac{c_{19}}{c_{4}}L^{-\eta},
	\label{sin/sin}
\end{eqnarray}
where in the final line we used a new assumption on the exponents
\begin{equation}
	\nu+\ep-1\le-\eta.
	\label{cond3}
\end{equation}

We now use the estimate (\ref{sin/sin}) and the approximate solution
(\ref{discap}) to write the sum (\ref{Sjq}) more explicitly as
\begin{equation}
	S_{j}^{(q)}
	=
	\frac{A^{2}}{\sin\zeta_{\hell_{q}}}
	\sum_{\ell\in\hI_{q}}\chi[j(\ell)=j]
	(\cos[Z_{\ell}+\xi])^{2}
	+R_{j}^{(q)},
	\label{Sjq2}
\end{equation}
where $R_{j}^{(q)}$ satisfies
\begin{equation}
	\abs{R_{j}^{(q)}}
	\le
	\frac{c_{20}A^{2}}{\sin\zeta_{\hell_{q}}}
	|\hI_{q}^{(j)}|L^{-\eta},
	\label{|R|<}
\end{equation}
where $\hI_{q}^{(j)}$ is the subset of $I_{q}$ with $j(\ell)=j$, and
$|\hI_{q}^{(j)}|$ denotes the number of its elements.

We now want to evaluate the sum over $\cos^{2}$ terms in (\ref{Sjq2}).
Let $\tzeta_{q}=\zeta_{\hell_{q}}$, and $\tZ_{q}=Z_{\hell_{q}-1}$.
Then we have
\begin{eqnarray}
	&&
	\abs{
	(\cos[Z_{\ell}+\xi])^{2}-
	(\cos[\tZ_{q}+\xi+\tzeta_{q}(\ell-\hell_{q})])^{2}
	}
	\ret&&
	\le
	2\max_{\ell\in\hI_{q}}
	\abs{Z_{\ell}-\cbk{\tZ_{q}+\tzeta_{q}(\ell-\hell_{q})}}
	\ret&&
	\le
	2\rbk{\max_{\ell\in\hI_{q}}|\zeta_{\ell}-\tzeta_{q}|}
	|\hI_{q}|
	\ret&&
  \le
	2\rbk{\max_{\ell\in\hI_{q}}(1-{\alpha_{\ell}}^{2})^{-1/2}}
	\abs{\alpha_{\ell}-\alpha_{\hell_{q}}}|\hI_{q}|
	\ret&&
	\le
	2\sqrt{\frac{g}{c_{4}}}(c_{19})^{2}L^{2\nu+(\ep/2)-1}
	\ret&&
	\le
	2\sqrt{\frac{g}{c_{4}}}(c_{19})^{2}L^{-\eta},
	\label{cos-cos}
\end{eqnarray}
where the final line again makes use of the new assumption
\begin{equation}
	2\nu+\frac{\ep}{2}-1\le\-\eta.
	\label{cond4}
\end{equation}

We introduce a new constant
$\txi_{q}=\tZ_{q}+\xi-\tzeta_{q}\hell_{q}$.
Then (\ref{cos-cos}) essentially means that
$(\cos[Z_{\ell}+\xi])^{2}\simeq
(\cos[\tzeta_{q}\ell+\txi_{q}])^{2}$.
So we are motivated study the sum
\begin{eqnarray}
	 &  &
	\sum_{\ell\in\hI_{q}}\chi[j(\ell)=j]
	(\cos[\tzeta_{q}\ell+\txi_{q}])^{2}
	\ret&&
	=\frac{|\hI_{q}^{(j)}|}{2}+
	\frac{1}{4}\sum_{\ell\in\hI_{q}}
	\chi[j(\ell)=j]
	\rbk{
	e^{2i(\tzeta_{q}\ell+\txi_{q})}+e^{-2i(\tzeta_{q}\ell+\txi_{q})}
	},
	\label{chicos}
\end{eqnarray}
which is a good approximation to the sum in (\ref{Sjq2}).
We expect the sum over oscillating exponential
in (\ref{chicos}) to be small, but this is
not straightforward.
Indeed, the sum is not at all small if a ``resonance'' between
$\chi[j(\ell)=j]$ and the exponential term takes place.
By using (\ref{ljk2}), we rewrite the sum of the first exponential
term (times $e^{-2i\txi_{q}}$) as
\begin{eqnarray}
	 &  &
	\sum_{\ell\in\hI_{q}}\chi[j(\ell)=j]e^{2i\tzeta_{q}\ell}
	\ret&&
	=
	\sum_{k{\rm s.t.}\ell(j,k)\in\hI_{q}}
	\exp\rbk{
	2i\tzeta_{q}\hell_{q}
	+2i\tzeta_{q}\sum_{j'=1}^{n}\sqbk{
	\frac{\tM_{j'}}{\tM_{j}}k+\eta_{j,j'}
	}}
	\ret&&
	=
	\sum_{p=0}^{\kmax-\kmin}
	\exp\rbk{
	2i\tzeta_{q}\hell_{q}
	+2i\tzeta_{q}\sum_{j'=1}^{n}\sqbk{
	\frac{\tM_{j'}}{\tM_{j}}p
	+\frac{\tM_{j'}}{\tM_{j}}\kmin
	+\eta_{j,j'}
	}},
	\label{chiexplong}
\end{eqnarray}
where $[\cdots]$ is the Gauss symbol.
The sum in the second line is over $k$ such that $\ell(j,k)\in\hI_{q}$
for the fixed $j$, and we denote this range of $k$ as
$\{\kmin,\ldots,\kmax\}$.
We let $\bas=[(\kmax-\kmin)/\tM_{j}]$, and write
$p=\tM_{j}s+r$.
Then the above sum becomes
\begin{eqnarray}
	 & = &
	e^{2i\tzeta_{q}\hell_{q}}
	\sum_{s=0}^{\bas-1}\sum_{r=0}^{\tM_{j}-1}
	\exp\rbk{
	2i\tzeta_{q}\sum_{j'=1}^{n}\sqbk{
	\frac{\tM_{j'}}{\tM_{j}}(\tM_{j}s+r)
	+\teta_{j,j'}
	}}
	\ret&&
	+e^{2i\tzeta_{q}\hell_{q}}
	\sum_{p=\bas\tM_{j}}^{\kmax-\kmin}
	\exp\rbk{
	2i\tzeta_{q}\sum_{j'=1}^{n}\sqbk{
	\frac{\tM_{j'}}{\tM_{j}}p+\teta_{j,j'}
	}}
	\ret
	&=&
	e^{2i\tzeta_{q}\hell_{q}}
	\sum_{s=0}^{\bas-1}\sum_{r=0}^{\tM_{j}-1}
	\exp\rbk{
	2i\tzeta_{q}\sum_{j'=1}^{n}\cbk{
	\tM_{j'}s+
	\sqbk{\frac{\tM_{j'}}{\tM_{j}}r+\teta_{j,j'}}
	}}
	\ret&&
	+(\mbox{the same second term})
	\ret
	&=&
	e^{2i\tzeta_{q}\hell_{q}}
	\frac{1-e^{2i\tzeta_{q}\tM\bas}}{1-e^{2i\tzeta_{q}\tM}}
	\sum_{r=0}^{\tM_{j}-1}
	\exp\rbk{
	2i\tzeta_{q}\sum_{j'=1}^{n}
	\sqbk{\frac{\tM_{j'}}{\tM_{j}}r+\teta_{j,j'}}
	}
	\ret&&
	+(\mbox{the same second term}),
	\label{chiexplonger}
\end{eqnarray}
where
\begin{equation}
	\tM=\sum_{j=1}^{n}\tM_{j}.
	\label{Mtil}
\end{equation}
By taking the absolute value of (\ref{chiexplong}) and
(\ref{chiexplonger}), we find
\begin{eqnarray}
	\abs{
	\sum_{\ell\in\hI_{q}}
	\chi[j(\ell)=j]e^{2i\tzeta_{q}\ell}
	}
	& \le &
	\rbk{\frac{|\sin\tzeta_{q}\tM\bas|}{|\sin\tzeta_{q}\tM|}+1}
	\tM_{j}
	\ret
	 & \le &
	\frac{2\tM}{|\sin\tzeta_{q}\tM|}.
	\label{chiexpabs}
\end{eqnarray}
As we have anticipated, the right-hand side of (\ref{chiexpabs}) is
usually small (compared with $|\hI_{q}^{(j)}|$),
but becomes large near the ``resonance'' points where
$\tzeta_{q}\tM$ is equal to an integer multiple of $2\pi$.

To control the final sum (\ref{Sj2}), we classify the subintervals
$\hI_{1},\ldots,\hI_{Q}$ into ``good'' ones and ``bad'' ones.
A subinterval $\hI_{q}$ is said to be good if
\begin{equation}
	\frac{2\tM}{|\sin\tzeta_{q}\tM|}
	\le
	c_{20}|\hI_{q}|L^{-\eta},
	\label{good}
\end{equation}
and to be bad otherwise.
We want to know the possible number of the bad subintervals.
First we note that the perfect resonance
$\zeta_{\ell}\tM=2\pi\times(\mbox{integer})$
can take place in the whole
interval $I$ at most once, since $\zeta_{\ell}$ varies at most by
$O(L^{\nu+(\ep/2)-1})$ within $I$.
We suppose that there happens to be a perfect resonance (with
$\zeta_{\rm res}$) within $I$, and see how many subintervals around
it are ``infected'' and become bad.
From (\ref{good}) and that
$|\hI_{q}|=[c_{19}L^{\nu}]$, we get
\begin{equation}
	|\tzeta_{q}-\zeta_{\rm res}|
	\ge
	\frac{L^{\eta-\nu}}{c_{20}c_{19}},
	\label{goodzeta}
\end{equation}
for $\tzeta_{q}$ corresponding to a good $\hI_{q}$.
This in turn means that the total number of the bad subintervals is
bounded from above by
\begin{eqnarray}
	\frac{L^{\eta-\nu}}{c_{20}c_{19}}
	\rbk{\max_{\ell,\ell'\in I}|\zeta_{\ell}-\zeta_{\ell'}|}^{-1}
	Q
	&\le&
	\frac{L^{\eta-\nu}}{c_{20}c_{19}}
	(gc_{5}L^{-2\theta})^{-1}Q
	\ret&&
	=
	\frac{1}{gc_{5}c_{20}c_{19}}L^{\eta-\nu+2\theta}Q
	\ret&&
	\le
	\frac{1}{gc_{5}c_{20}c_{19}}L^{-\eta}Q,
	\label{badnum}
\end{eqnarray}
where in the final line we assumed that
\begin{equation}
	\eta-\nu+2\theta\le-\eta.
	\label{cond5}
\end{equation}

To summarize, we have proved the following for the desired sum
$S_{j}^{(q)}$ (\ref{Sjq}).
When $\hI_{q}$ is a ``good'' subinterval, we have
\begin{equation}
	\abs{
	S_{j}^{(q)}-
	\frac{A^{2}}{2\sin\zeta_{\hell_{q}}}|\hI_{q}^{(j)}|
	}
	\le
	c_{21}\frac{A^{2}}{\sin\zeta_{\hell_{q}}}|\hI_{q}^{(j)}|
	L^{-\eta}.
	\label{Sjiq<}
\end{equation}
When $\hI_{q}$ is a ``bad'' subinterval, we have essentially no
control on $S_{j}^{(q)}$, and can only say
\begin{equation}
	\abs{S_{j}^{(q)}}
	\le
	c_{22}\frac{A^{2}}{\sin\zeta_{\hell_{q}}}|\hI_{q}^{(j)}|,
	\label{Sjq<2}
\end{equation}
but we have the estimate (\ref{badnum}) for the possible number of
the ``bad'' intervals.
Recalling (\ref{Sj2}), we sum up (\ref{Sjiq<}) and (\ref{Sjq<2})
with (\ref{badnum}) in mind.
We then get
\begin{equation}
	\abs{S_{j}-\tilde{S}_{j}}
	\le
	c_{23}\tilde{S}_{j}L^{-\eta},
	\label{Sj-Stj}
\end{equation}
with
\begin{equation}
	\tilde{S}_{j}=\sum_{q=1}^{Q}
	\frac{A^{2}}{2\sin\zeta_{\hell_{q}}}|\hI_{q}^{(j)}|.
	\label{Stj}
\end{equation}
This immediately leads us to our goal (\ref{partB}) that
\begin{equation}
	\abs{\frac{S_{j}}{\sum_{j'=1}^{n}S_{j'}}-W_{j}}
	\le
	c_{1}W_{j}L^{-\eta},
	\label{partBpart2}
\end{equation}
with
\begin{equation}
	W_{j}=
	\frac{|\hI_{q}^{(j)}|}{\sum_{j'=1}^{n}|\hI_{q}^{(j')}|},
	\label{Wjpart2}
\end{equation}
which is independent of $q$ and is equal to (\ref{Wwj}).

\subsection{Boltzmann factor from regions with long wave length}
\label{s:long}
We still have to prove the main Lemma~\ref{l:part} for the interval
$I_{\omega}$ with $\omega=1,\ldots,\Gamma$ and
$\omega=\Omega-\Gamma+1,\ldots,\Omega$.
In these intervals, the wave length of $\phi_{\ell}$ is quite large,
and the resonance effect (which made the estimates in
Section~\ref{s:res} difficult) does not take place.
The proof is rather straightforward, and we will be sketchy here.

In these intervals, we use one of the Lemmas~\ref{l:discap},
\ref{l:contap}, or \ref{l:contap2} to get controlled approximations
for the solution $\phi_{\ell}$ of the \Sch\ equation (\ref{Sch}).
In most of the situations, $|\phi_{\ell}|$ can be treated essentially
as a constant, and the estimate of (\ref{Sj}) becomes as trivial as
\begin{equation}
	S_{j}
	=
	\sum_{\ell\in I}\chi[j(\ell)=j]|\phi_{\ell}|^{2}
	\simeq
	|\phi_{\ell}|^{2}\sum_{\ell\in I}\chi[j(\ell)=j].
	\label{Sjeasy}
\end{equation}
The worst case that we have to worry about is when $\phi_{\ell}$
changes its sign inside $I$.
To investigate such a situation, we approximate
$\phi_{\ell}\simeq(\ell-\ell_{\rm min})$, and consider a small
interval $I=\{\ell_{\rm min},\ldots,\ell_{\rm max}\}$.
The sum to be evaluated is
\begin{equation}
	S_{j}=\sum_{\ell=\ell_{\rm min}}^{\ell_{\rm max}}
	\chi[j(\ell)=j](\ell-\ell_{\rm min})^{2}.
	\label{Sjlin}
\end{equation}
Though $j(\ell)$ may be a very complicated function of $\ell$,
(\ref{ljk2}) implies that it is (at least) a periodic function of
$\ell$ with the period $\tM=\sum_{j=1}^{n}\tM_{j}$.
Assuming $\ell_{\rm max}-\ell_{\rm min}=\bas\tM$, we evaluate for
$r<\tM$ the sum
\begin{eqnarray}
	\sum_{s=0}^{\bas}(s\tM+r)^{2}
	& = &
	\frac{\bas(2\bas+1)(\bas+1)}{6}\tM^{2}
	+\bas(\bas+1)\tM r+(\bas+1)r^{2}
	\ret
	 & = &
	\frac{\bas(2\bas+1)(\bas+1)}{6}\tM^{2}(1+O(1/\bas)),
	\label{bossum}
\end{eqnarray}
to see that the (unwanted)
$r$-dependent terms are of order $O(1/\bas)$.
Thus in order for $S_{j}/\sum_{j'=1}^{n}S_{j'}$ to become the desired
Boltzmann factor with relative error $O(L^{-\eta})$, we must have
that $\bas\ge O(L^{\eta})$, and hence
\begin{equation}
	\ell_{\rm max}-\ell_{\rm min}\ge {\rm const.}L^{\eta}.
	\label{lmax-lmin}
\end{equation}
In the present situation, $\ell_{\rm max}-\ell_{\rm min}$ is
determined by (the smaller of) the length of the interval and the wave
length of the oscillation of $\phi_{\ell}$.
Since the wave length in this region is longer than a constant times
$L^{\ep/2}$, the required conditions for the exponents
are
\begin{equation}
	\frac{\ep}{2}\ge\eta,
	\quad
	\frac{1}{3}-\eta\ge\eta.
	\label{cond6}
\end{equation}
These guarantee the Lemma~\ref{l:part} for the desired intervals.

\subsection{Determination of the exponents}
\label{s:exp}
In order to complete the lengthy proof of Lemma~\ref{l:part}, we have
to fix the values of the exponents $\eta$, $\ep$, $\theta$ and $\nu$
so that several requirements are satisfied.

We now recall the requirements about  the exponents appeared as
(\ref{cond0}), (\ref{cond0.5}), (\ref{cond0}), (\ref{cond1}),
(\ref{cond2}), (\ref{cond3}), (\ref{cond4}), (\ref{cond5}),
and (\ref{cond6}), which are
\begin{eqnarray}
	1-\ep & > & \frac{1}{3}-\eta,
	\ret
	-2\theta+\ep & \le & 0,
	\ret
	\ep & \le & \frac{1}{2},
	\ret
	-2\theta+\frac{\ep}{2} & \le & -\eta,
	\ret
	\nu+\ep-1 & \le & -\eta,
	\ret
	2\nu+\frac{\ep}{2}-1 & \le & -\eta,
	\ret
	\eta-\nu+2\theta & \le & -\eta,
	\ret
	\frac{\ep}{2}& \ge & \eta,
	\ret
	\frac{1}{3}-\eta & \ge & \eta.
	\label{exponents}
\end{eqnarray}
As a solution which satisfies all of (\ref{exponents}), we shall
choose
\begin{eqnarray}
	\eta=\theta=\frac{1}{12},\quad
	\nu=\frac{1}{3},\quad
	\ep=\frac{1}{6}.
	\label{theexponents}
\end{eqnarray}
This completes the proof.

\section{General upper bounds for $|\varphi_{(j,k)}|$}
\label{s:gen}
In the main paper, we noted that the ``hypothesis of equal weights for
eigenstates'' can be {\em partially} proved.
Let me describe precise statements.

Recall that the ``hypothesis of equal weights for
eigenstates'' consists of two parts;
\begin{enumerate}
\item
$|\varphi_{(j,k)}|^{2}$ is negligible in the ``classically inaccessible
region.''
\item
$|\varphi_{(j,k)}|^{2}$ takes appreciable values all over in the
``classically accessible region'', and the value is essentially
determined by a function $f$ of the energy
$E-(\varepsilon_{j}+B_{k})$.
\end{enumerate}
Clearly the second point is much more subtle.
In fact if we take models with certain conservation laws, the second
point is easily violated.
The main point of our (unproven) hypothesis is that the above 2 holds
in a general model which do not have any special symmetries or
conservation laws.

On the other hand the first point is more universal, and may be
stated for a large class of models rather easily.
Here we prove two results which justify this second point.
There can be various theorems under different assumptions.
We here present two simple
theorems, which are in some sense
complimentary with each other.

We use the same notation as in the main paper, but now take
general
coupling Hamiltonian $H'$, and
denote its matrix elements as
\begin{equation}
	V_{j,k;j',k'}
	=
	\bkt{\Psi_{j}\otimes\Gamma_{k},
	H'
	\Psi_{j'}\otimes\Gamma_{k'}
  }.
	\label{V}
\end{equation}
We also denote by $\HiS$ and $\HiB$ the Hilbert spaces for the
subsystem and the bath.

As in the main paper, the the Schr\"{o}dinger equation is
\begin{equation}
	(H_{\rm S}\otimes{\bf 1}_{\rm B}
	+{\bf 1}_{\rm S}\otimes H_{\rm B}+H')\Phi
	=
	E\Phi.
	\label{Xi1}
\end{equation}
Again expanding the eigenstate as
\begin{equation}
	\Phi=\sum_{j=1}^{n}\sum_{k=1}^{N}
	\varphi_{(j,k)}\Psi_{j}\otimes\Gamma_{k},
	\label{Xi2}
\end{equation}
 (\ref{Xi1}) is rewritten as
\begin{equation}
	\cbk{E-(\ep_{j}+E_{k})}\phi_{(j,k)}
	=
	\sum_{j',k'}
	V_{j,k;j',k'}\,\phi_{(j',k')},
	\label{Sch0}
\end{equation}
for any $j=1,\ldots,n$ and $k=1,\ldots,N$.

We define
\begin{equation}
	\lambda
	=
	\sup_{j,k}
	\sum_{j',k'}\abs{V_{j,k;j',k'}}.
	\label{Vbar}
\end{equation}

The first result is very simple.
\begin{theorem}
Let a normalized state $\Phi\in\HiS\otimes\HiB$ satisfy (\ref{Xi1}).
Then the coefficients $\phi_{(j,k)}$ defined by (\ref{Xi2}) satisfy
\begin{equation}
	|\phi_{(j,k)}|\le
	\frac{\lambda}{\abs{E-(\ep_{j}+B_{k})}}.
	\label{xi1}
\end{equation}
\end{theorem}
\proof
From (\ref{Sch0}), we see that
\begin{eqnarray}
	|\phi_{(j,k)}|
	&\le&
	\frac{
	\sum_{j',k'}|V_{j,k;j',k'}|\,|\phi_{(j',k')}|
	}
	{\abs{E-(\ep_{j}+B_{k})}}
	\ret
	&\le&
	\frac{\lambda}
	{\abs{E-(\ep_{j}+B_{k})}},
	\label{xibound}
\end{eqnarray}
where we noted $|\phi_{(j,k)}|\le1$ because $\Phi$ is normalized.\qed

The bound (\ref{xibound}) is very crude, but gives us the idea that
$|\phi_{(j,k)}|$ may be large for
$|E-(\ep_{j}+B_{k})|\lesssim\lambda$,
and is small for $|E-(\ep_{j}+B_{k})|\gg\lambda$.

To get stronger result, we further assume that there is a constant
$D>0$ such that $V_{j,k;j',k'}=0$ if
$|(\ep_{j}+B_{k})-(\ep_{j'}+B_{k'})|>D$.
In other words, the interaction Hamiltonian has no matrix elements
between the basis
states whose unperturbed energies differ by more than $D$.
The assumption may hold for some interaction Hamiltonians which
represent (near) elastic scattering, but not for some interactions.
Under this assumption, we have the following.
\begin{theorem}
Let a normalized state $\Phi\in\HiS\otimes\HiB$ satisfy (\ref{Xi1}).
Then the coefficients $\phi_{(j,k)}$ defined by (\ref{Xi2}) satisfy
\begin{equation}
	|\phi_{(j,k)}|\le
	\frac{1}{n(j,k)!}
	\rbk{
	\frac{\lambda}{D}
	}^{n(j,k)},
	\label{xi2}
\end{equation}
where the integer $n(j,k)$ is defined as
\begin{equation}
	n(j,k)
	=
	\sqbk{
	\frac{\abs{E-(\ep_{j}+B_{k})}}
	{D}
	}.
	\label{njl}
\end{equation}
Here $[\cdots]$ is the Gauss symbol.
\end{theorem}
\proof
The bound (\ref{xi2}) (with the convention $0!=1$) is trivial for
$j$, $k$ such that $n(j,k)=0$.
Assume (\ref{xi2}) for all $j'$, $k'$ such that
$n(j',k')\le n-1$.
Take $j$, $k$ such that $n(j,k)=n$.
Then from (\ref{Sch}) we have
\begin{eqnarray}
	|\phi_{(j,k)}|
	&\le&
	\frac{\lambda}{\abs{E-(\ep_{j}+B_{k})}}
	\max_{
	(j',k') {\rm s.t.}
	|(\ep_{j}+B_{k})-(\ep_{j'}+B_{k'})|\le D
	}
	|\phi_{(j',k')}|
	\ret
	&\le&
	\frac{\lambda}{nD}
	\frac{1}{(n-1)!}
	\rbk{\frac{\lambda}{D}}^{(n-1)}
	\ret
	&\le&
	\frac{1}{n!}
	\rbk{\frac{\lambda}{D}}^n,
	\label{xibound2}
\end{eqnarray}
which proves the desired bound.\qed

\bigskip
The bound (\ref{xi2}) shows that $|\phi_{(j,k)}|$ actually decays very
rapidly in the ``classically inaccessible region.''
\end{document}